\documentclass[11pt]{article}
\usepackage[T1]{fontenc}
\usepackage{geometry}
\usepackage{natbib}

\usepackage[figuresright]{rotating}

\usepackage{setspace}

\usepackage{amsmath}
\usepackage{amsfonts}
\usepackage{babel}
\usepackage{bbm}
\usepackage{graphicx}
\usepackage{subcaption}
\usepackage{float}
\usepackage{url}
\usepackage{color}
\usepackage[dvipsnames]{xcolor}
\usepackage{multicol,multirow}
\usepackage{lineno}

\usepackage[section]{placeins}

\usepackage{tikz}
\usepackage{etoolbox}
\newrobustcmd*{\mytriangle}[1]{\tikz{\filldraw[draw=#1,fill=#1] (0,0) --
(0.2cm,0) -- (0.1cm,0.2cm);}}

\usepackage{algorithm}
\usepackage[noend]{algpseudocode}

\usepackage{setspace}

\usepackage{authblk}

\title{Inferring HIV Transmission Patterns from Viral Deep-Sequence Data via Latent Typed Point Processes}
\author[1$\dagger$]{Fan Bu}
\author[2]{Joseph Kagaayi}
\author[3]{Kate Grabowski}
\author[4]{Oliver Ratmann}
\author[5]{Jason Xu}

\affil[1]{Department of Biostatistics, University of California - Los Angeles, Los Angeles, CA 90024, USA}
\affil[2]{School of Public Health, Makerere University, Kampala, Uganda}
\affil[3]{School of Medicine, Johns Hopkins University, Baltimore, MD 21218, USA}
\affil[4]{Department of Mathematics, Imperial College London, London SW7 2AZ, UK}
\affil[5]{Department of Statistical Science, Duke University, Durham, NC 27708, USA}
\affil[$^\dagger$]{On behalf of the Rakai Community Cohort Study and PANGEA-HIV}

\date{}

\begin{document}

\setstretch{1.5}

\maketitle

\begin{abstract}
Viral deep-sequencing data play a crucial role toward understanding disease transmission network flows, because the higher resolution of these data compared to standard Sanger sequencing provide evidence into the direction of infectious disease transmission. To more fully utilize these rich data and account for the uncertainties in phylogenetic analysis outcomes,  we propose a spatial Poisson process model to uncover HIV transmission flow patterns at the population level. We represent pairings of two individuals with viral sequence data as typed points, with coordinates representing covariates such as gender and age, and the point type representing the unobserved transmission statuses (linkage and direction). Points are associated with observed scores on the strength of evidence for each transmission status that are obtained through standard deep-sequenece phylogenetic analysis. Our method is able to jointly infer the latent transmission statuses for all pairings and the transmission flow surface on the source-recipient covariate space. In contrast to existing methods, our framework does not require  pre-classification of the transmission statuses of data points, instead learning them probabilistically through a fully Bayesian inference scheme. By directly modeling continuous spatial processes with smooth densities, our method enjoys significant computational advantages compared to previous methods that rely on discretization of the covariate space. We demonstrate that our framework can capture age structures in HIV transmission at high resolution, and bring valuable insights in a case study on viral deep-sequencing data from Southern Uganda.
\end{abstract}

%
%


\maketitle

\section{Introduction}
\label{sec:introduction}

As a decades-long global pandemic, the human immunodeficiency virus (HIV) has most severely affected Africa with 1 in every 25 adults living with the HIV virus, accounting for more than two-thirds of infections worldwide~\citep{eisinger2018ending,fauci2020four}. 
International public health organizations have proposed to control the HIV epidemic by targeting intervention efforts towards  populations most at risk of HIV acquisition and transmission~\citep{glynn2001young,pettifor2008keep,karim2010preventing,jewkes2010intimate, saul2018determined}. To achieve this, it is important to understand and characterize transmission patterns between different groups and sub-populations \citep{wilson2008know}. 

To this end, this article proposes novel methodology to infer the relative transmission flows between different groups of individuals. Throughout this paper we will focus on heterosexual transmission flows across age groups. We model the underlying ground truth transmission flows as latent surfaces in a plane whose axes denote respectively the ages of the source and the recipient. A transmission pair is thus represented by a point on this surface with coordinates representing the phylogenetically likely source and recipient ages. The population of all transmission pairs then corresponds to a realized point pattern from the underlying flow surface. We expect that population groups with similar age attributes behave similarly, and so, in statistical terms, the underlying transmission flows can be modelled in the same way as continuous latent spatial surfaces on a compact domain specified by longitudes and latitudes~\citep{ji2009spatial, kutoyants2012statistical}. The key scientific challenge, however, arises due to the unobserved transmission pathways in the phylogenetically likely transmission pairs---there is uncertainty about whether or not there is a transmission link, and in which direction transmission takes. 
Addressing the question ``who infected whom?'' is fundamental for inferring the population-level transmission flow. 

In order to reliably infer transmission pathways between individuals, we leverage the outputs from modern phylogenetic and phylodynamic analyses using viral deep-sequencing data. Recently developed viral deep-sequencing pipelines have made it possible to estimate the linkage and direction of transmission between infected individuals~\citep{Severson2016Phylo,leitner2018phylogenetic,wymant2018phyloscanner,ratmann2019inferring}. Multiple viral variants are observed for each individual and phylogeographic techniques can then be applied at the individual level to produce evidence about the evolutionary relationships between individual viral sequences, which inform the transmission relationships between individuals ~\citep{wymant2018phyloscanner,zhang2021evaluation}. The outputs from these phylogenetic analysis pipelines typically take the form of two summary scores that represent (a) how likely deep-sequenced individuals shared a transmission link and (b) how probable transmission occurred in a specific direction. These scores can provide rich information about transmission flows between groups of individuals, for example between locations \citep{ratmann2020quantifying} or between age and gender groups \citep{bbosa2020phylogenetic,xi2021inferring, hall2021demographic}. Importantly, the phylogenetic summary scores are imprecise at the level of two individuals and cannot be used to ``prove" transmission between two individuals, because the phylogenetic signal is asssociated with measurement error and not fully consistent for the same individuals across independent evaluations on different parts of the genome~\citep{ratmann2019inferring,hall2021demographic,zhang2021evaluation}. In addition, phylogenetic analysis typically only outputs the ``maximum likelihood'' phylogenetic structures of viral sequences, without reporting the corresponding uncertainties for alternative structures or relationships. 
In practice, existing approaches for inferring population-level transmission flows neglect such uncertainty by arbitrarily thresholding the phylogenetic summary scores (e.g., by only keeping highly likely transmission pairs with likelihood scores for transmission linkage $>0.6$ and transmission direction $>0.5$, commonly called ``source-recipient pairs"). As a result, the varying strengths of phylogenetic evidence are not accounted for in existing approaches. Further, a large fraction of data points are not even considered in analysis because among all pairs with some evidence of transmission, only the subset of highly likely source-recipient pairs are classified as ``source-recipient pairs'' and included for study. In short, there is a substantial methodological gap in existing approaches for utilizing phylogenetic evidence in that (1) differential evidence confidence is neglected and (2) a large amount of data must be discarded.

This paper addresses these limitations  by 
proposing a coherent statistical model which jointly learns from demographic information (such as gender and age) and phylogenetic evidence. We introduce a marked latent spatial process model on the age space, where each transmission pair of individuals (represented by their paired ages as coordinates on the space) is associated with  ``marks'' that contain the phylogenetic summary scores. That is, each potential transmission pair is assigned a latent ``type" random variable that indicates the unknown transmission statuses (linkage and direction). The distribution of transmission flow between age groups and the distribution of the ``marks'' (phylogenetic scores) both depend on the latent type variable. We derive the likelihood of the complete model, so that basing inference in a data-augmented Bayesian framework then allows us to probabilistically learn the latent type for each potential transmission pair. In particular, this allows the evidence strength for each pair of infected individuals to be quantified--- for example, we may conclude a 85\% posterior probability for a specific heterosexual pair to be linked through transmission, and this pair would contribute more to the learning of transmission flow surfaces compared to another pair with a 40\% posterior probability of linkage. More importantly, this joint modeling approach enables us to make use of substantially more data, as ``low-confidence'' pairs with lower phylogenetic scores reflecting weaker evidence for transmission linkage or direction are downweighted in a data-driven manner rather than discarded.

Not only will the latent spatial point process approach leverage more evidence with uncertainty quantification, but it also admits a more computationally efficient solution. This is largely due to the continuous formulation of the transmission flow age space. A common approach in past epidemiological studies makes use of discrete grids based on pre-specified age groups (such as 1-year or 5-year age groups), but these heuristic groupings can lead to computationally intensive downstream analysis~\citep{hyman1994threshold, heuveline2004impact, sharrow2014modeling}. For instance, recent work of~\citet{xi2021inferring} introduces a semi-parametric Poisson model for flow counts on discrete age strata and other demographic attributes. The number of observed points is typically considerably smaller than the number of all cells in the high-dimensional age grid; those structural zeros in discrete age cells require a considerable amount of book-keeping as well as heavy computation for model smoothing. Instead, point pattern approaches with a continuous underlying flow surface can be at once more general and more computationally efficient, even with the addition of latent transmission types. These advances free us from specifying ad hoc thresholds on the phylogenetic summary scores, and in turn allow the larger set of available data points with lower evidence confidence to be considered, with each data point weighed according to its phylogenetic evidence. Interestingly, we will also show that the point process model borrows information in a two-way manner, leveraging the additional data to learn the transmission flows, and using the transmission flows to learn the point types. Although our statistical methodology is motivated by applications in HIV transmissions, it represents a general framework that can be applied to studying transmission dynamics for many other infectious diseases such as hepatitis C virus, human papillomavirus and Monkeypox virus.

\paragraph{Prior work} 
Spatial Poisson process models have been widely applied to the study of point-referenced two-dimensional data~\citep{banerjee2003hierarchical,huber2011spatial,cressie2015spatial}. Heterogeneity in spatial point patterns is often modelled through non-homogeneous Poisson processes (NHPPs), where the structure of the intensity function can be described using various choices of Bayesian mixture models. NHPP intensity functions have been parameterized through Markov random fields for piecewise constant functions based on Voronoi tessellations \citep{heikkinen1998non}, weighted Gamma process mixtures of non-negative kernel functions~\citep{lo1989class,wolpert1998poisson,ishwaran2004computational}, Gaussian process mixtures of log-transformed  components~\citep{moller1998log,brix2001spatiotemporal,adams2009tractable}, and Dirichlet process mixtures~\citep{ji2009spatial,zhou2015spatio,zhao2021modelling}. \cite{kim2022erlang} offer an excellent summary. Our framework builds on previous developments in Dirichlet process mixtures that focus on learning a normalized functional form of the NHPP intensity~\citep{kottas2007bayesian,kottas2008modeling,taddy2012mixture}, which admits tractable and efficient inference procedures and in this application corresponds to the transmission flows that we seek to infer. 

In recent years, Poisson process models have been extended to study spatial point patterns that are latent or partially observed, in that only certain indirect ``signals'' associated with the underlying spatial pattern are observed, with uncertainty about the quantity and locations of the latent points~\citep{vedel2007spatio,ji2009spatial}. Here, a ``signal'' is any additional information associated with a data point in the spatial pattern, which may depend on the location of the associated spatial point. The signal can be the height of trees in a forest survey, temperature of water sample in a lake, or in our application, the phylogenetic evidence between a pair of infected individuals. Joint modelling of the signals and the latent point process through Bayesian data augmentation has been successful, thanks to the ease of incorporating missing information as latent variables in a Bayesian inference framework~\citep{givens1997publication}. However, to our knowledge, much of the existing work in spatial Point process models focuses on \emph{one} set or type of spatial points, rather than a combination of multiple types of latent point patterns whose ``types'' are associated with practical interpretation, but are not observed. Our framework also bridges this methodological gap by exploiting additional signals that inform the latent type labels in a marked spatial Poisson process model.

This paper is structured as follows. We first provide an overview of the motivating data and develop the model framework in Section~\ref{sec: data-model}. The latent Poisson spatial process model and a likelihood-based inference scheme are presented in Section~\ref{sec: inference}. We then investigate the accuracy of inference under the proposed model and its ability to differentiate between competing transmission flow hypotheses on simulate data in Section~\ref{sec:simulation-experiments}. Next, we illustrate the efficacy of the proposed approach on demographic and HIV deep-sequence data from the Rakai Community Cohort Study in Southern Uganda in Section~\ref{sec: case-study}. Finally, we discuss the merits and limitations of the proposed statistical model in Section~\ref{sec: conclusion}.

\section{Data and Model}
\label{sec: data-model}

\begin{figure}
    \centering
    \includegraphics[width=0.95\textwidth]{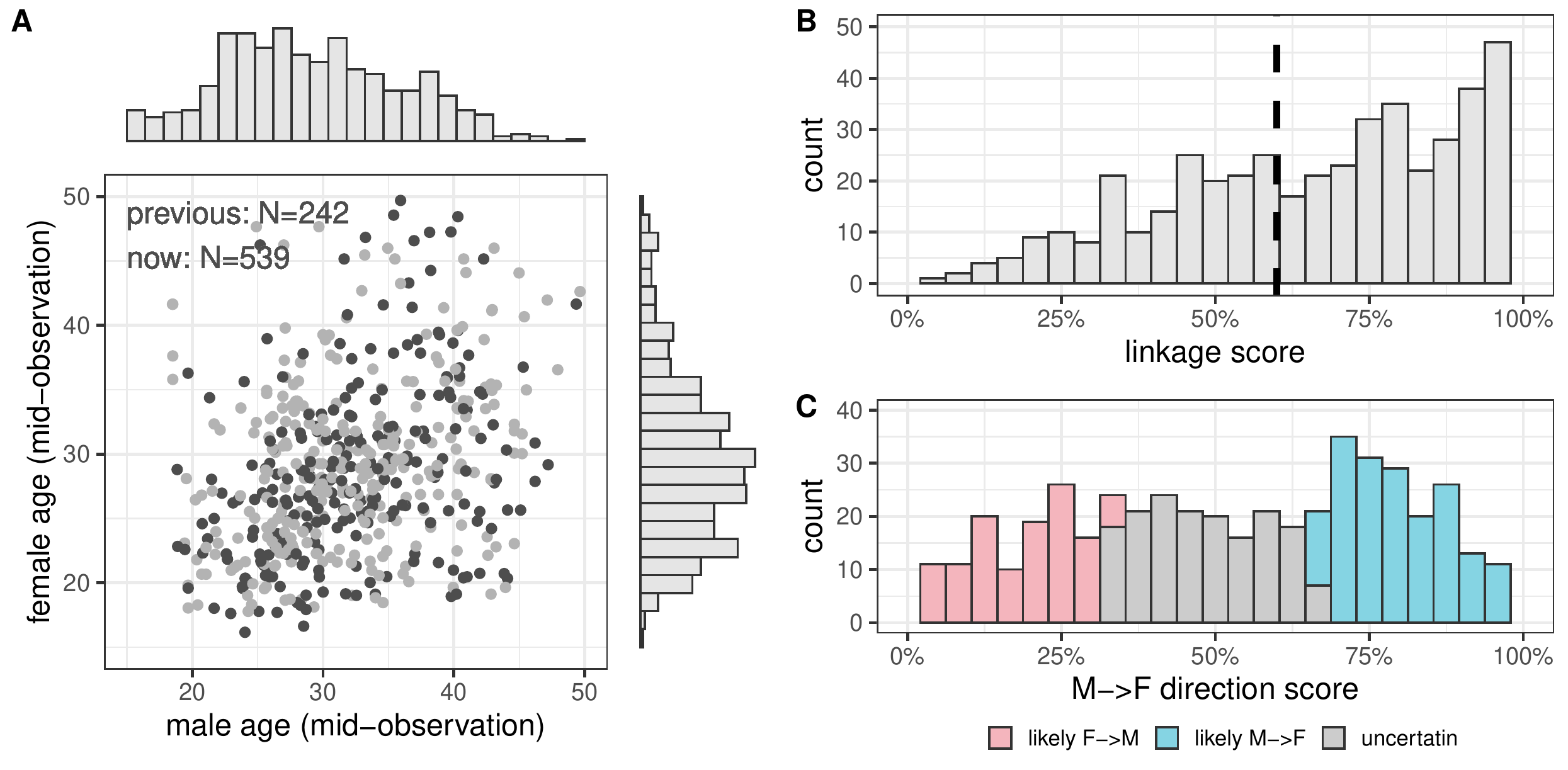}
    \caption{{\bf Point data and marks representing the output of HIV phylogenetic deep-sequence analysis on the sequence sample from Rakai, Uganda from 2010 to 2015 .} ({\bf A}) Paired ages of 539 heterosexual individuals who were inferred to be phylogenetically closely related with the HIV phylogenetic deep-sequence analysis using the \emph{phyloscanner} software on HIV deep-sequence data from 2652 study participants of the Rakai Community Cohort Study in Southern Uganda, 2010 to 2015. The age of the individuals in the closely related pairs was calculated at the midpoint of the observation period, and the age of the men and of the women are shown on the x-axis and y-axis respectively. Each data point is associated with two phylogenetic deep-sequence summary statistics in $[0,1]$, the linkage score ($\ell_i$) and the direction score ($d_i$) (see text). Points associated with high linkage and direction scores ($\ell_i \geq 0.6$ and $d_i \geq 0.67$) are shown in dark grey, and all other points are shown in light grey. Marginal histograms on the age of men and women are shown for all points. The typed point process model that we develop here aims to infer transmission flows using all data points rather than the highly likely ``source-recipient" pairs shown in dark grey. ({\bf B}) Histogram of the linkage scores across all data points. ({\bf C}) Histogram of the direction scores across all data points. Direction scores $d_i \leq 1/3$ indicate high confidence in female-to-male transmission (shown in red), and direction scores $d_i \geq 2/3$ indicate high confidence in male-to-female transmission (shown in blue).}
    \label{fig:ageScater_scoreHistogram}
\end{figure}

\subsection{Demographic and viral phylogenetic data of HIV infected individuals}

HIV deep-sequence data were obtained from blood samples of study participants living with HIV in the Rakai Community Cohort Study (RCCS), a longitudinal, open, population-based census and cohort study in southern Uganda~\citep{grabowski2017hiv}. Samples were collected between August 2011 and January 2015 from 5142 participants~\citep{ratmann2019inferring}, of whom 2652 had an HIV viral load $>1000$ copies/ml and viral sequence read depth and length sufficient for deep sequence data generation and analysis~\citep{wymant2018phyloscanner,ratmann2019inferring}. Detailed demographic, behavioural and healthcare data were collected for all participants, including their gender and age, the latter determined from self-reported birth date~\citep{grabowski2017hiv}.

The data that motivate our model consist of a subset of 539 heterosexual pairs of HIV infected RCCS participants considered as phylogenetically 
likely HIV transmission pairs. 
These likely transmission pairs were identified based on demographic survey responses about long-term sexual relationships as well as phylogenetic evidence extracted using the \emph{phyloscanner} deep-sequence analysis pipeline~\citep{wymant2018phyloscanner}. We provide an illustration of the data in Figure~\ref{fig:ageScater_scoreHistogram}. There are two main facets of the data. The first facet comprises the age of the two individuals in a pair at the midpoint of the observation period, which we denote by $\mathbb{S} = \{\mathbf{s}_i = (a_{i1},a_{i2})^T\}_{i=1}^N$, where the 2-dimensional vector $(a_{i1},a_{i2})$ records the male's age $a_{i1}$ and female's age $a_{i2}$ in the $i$th pair. In our application, the sample size is $N=539$. Our model will envision these points of paired ages as an observation from a spatial process describing the transmission structure. The second facet consists of two scores in the range of $(0,1)$ that are outputs from deep-sequence phylogenetic analyses of HIV deep-sequencing data with \emph{phyloscanner}. For each pair $i$ of individuals, \emph{phyloscanner} produces two scores --- a linkage score and a direction score --- by assessing the viral phylogenetic relationship of individuals in terms of the patristic distances and topological configurations of the viral reads in deep-sequence phylogenies, and then counting the observed patterns over sliding, overlapping genomic windows across the HIV genome~\citep{wymant2018phyloscanner,ratmann2019inferring}. The linkage score $\ell_i$ represents the posterior probability of the pair sharing a transmission link in the transmission process under a Binomial count model of window-specific linkage classifications \citep{ratmann2019inferring}. The direction score $d_i$, on the other hand, measures the posterior probability of transmission taking place from the male to the female in this pair under a similar count model. We collectively denote the phylogenetic scores for the $i$th pair by $\mathbf{x}_i = (\ell_i, d_i)^T$, and for brevity refer to the phylogenetic data as the ``marks" associated with each of the points $\mathbf{s}_i$ for $i = 1, \dotsc, N$. 

The key data challenge is the unobserved transmission relationship between pairs of HIV-infected individuals. Even for a pair with phylogenetic evidence suggesting high probability to be linked through disease transmission (with a high $\ell_i$ score), we do not have direct knowledge about the transmission linkage and direction. Our model (described later) will, therefore, probabilistically characterize the likelihoods of pairwise transmission linkage and direction in a data-driven manner.

\subsection{Substantial data information loss in existing modeling approaches}
Existing analyses attempt to address the unobserved pairwise transmission relationships by pre-classifying data points using heuristic thresholds on the phylogenetic summary scores \citep{xi2021inferring, ratmann2020quantifying,hall2021demographic}. For example, a potential transmission pair $i$ would be classified to represent a male-to-female transmission event if $\ell_i > 0.6$ \emph{and} $d_i > 0.67$; similarly, another pair $j$ would be taken as a female-to-male transmission if $\ell_j > 0.6$ \emph{and} $d_j < 0.33$~\citep{xi2021inferring}. Such thresholding \textit{a priori}, albeit procedurally simple, excludes a substantial proportion of data from the analysis, as  ``low-confidence'' pairs are completely discarded. Graphically, Figure~\ref{fig:ageScater_scoreHistogram} shows that all data points with linkage scores falling to the left of the vertical line in panel \textbf{B} and all data with direction scores in the gray region of panel \textbf{C} would be discarded, resulting in only 242 out of 539 total data points (see panel \textbf{A}) retained for analysis. 
%
Further, such pre-classification neglects differential strengths of phylogenetic evidence from data points classified as transmission events. Intuitively, we should have higher confidence about a likely transmission pair $i$ with $\ell_i = 0.98$ to represent a transmission event, compared to another pair $j$ with only $\ell_j = 0.61$. However, this difference in evidence strength is not reflected in existing methodology. Instead, all data points that are believed to be ``high confidence'' pairs are treated as equal and exchangeable, which fails to fully exploit the rich information summarized by the phylogenetic scores. In the next section, we introduce our modeling framework that makes better use of the data by jointly leveraging phylogenetic evidence and demographic information.

\subsection{The typed point process model}

We now specify a general framework for inferring population-level transmission flows from the point pattern $\mathbb{S} = \{\mathbf{s}_i = (a_{i1},a_{i2})^T\}_{i=1}^N$ with associated marks $\mathbf{x}_i = (\ell_i, d_i)^T$,~i.~e. the phylogenetic transmission and direction scores. The marks reflect the strength of phylogenetic evidence regarding the unknown ground truth transmission relationship in each heterosexual pair. For this purpose we introduce for each point a categorical latent random variable $c_i$ that encodes three possible events: transmission did not occur between the two individuals that define the point (denoted by $c_i=0$), transmission occurred from the male to the female individual (denoted by $c_i=1$), or transmission occurred from the female to the male individual (denoted by $c_i=-1$). For brevity, we refer to the $c_i$ as the latent ``types" associated with each observed point $\mathbf{s}_i$ and observed mark $\mathbf{x}_i$.
Intuitively, our modelling framework can then be thought of as a typed spatial Poisson process in the sense that the intensity function of the process and the distribution of the marks both depend on the event type, and then the typed process provides a generative model for marked point patterns. This framework, though motivated by our particular application, can be applied to many similar data sets of spatial point patterns of unknown types that are informed by marks associated with each point.

With this context in place, we begin by considering a spatial point pattern defined on a 2-dimensional space $\mathcal{S} \times \mathcal{S}$, where $\mathbb{S} = \{\mathbf{s}_i\}_{i=1}^N$ is the set of all points, and each point $\mathbf{s}_i = (s_{i1}, s_{i2})^T$ is represented as a point in this plane. We model the observed points in $\mathbb{S}$ as a realization of a 2D Poisson process on $\mathcal{S} \times \mathcal{S}$:
\begin{equation}
    \label{eq:all-PP}
    \mathbb{S} \sim PP(\boldsymbol\lambda).
\end{equation}
Following \cite{kottas2007bayesian} we decompose the intensity function $\boldsymbol\lambda$ into a scale component $\gamma$ and a density function $f(\cdot)$, 
\begin{equation}
    \label{eq:decompose-all}
    \boldsymbol\lambda(\cdot) = \gamma f(\cdot),
\end{equation}
so that $f(\cdot)$ satisfies $\int_{\mathcal{S}\times \mathcal{S}} f(s_1,s_2) ds_1ds_2 = 1$. This decomposition separates the intensity function into two terms which are simpler to write out in the likelihood function and make inference computationally tractable. We next model the density function $f(\cdot)$ as a mixture of $K$  different types:
\begin{equation}
\label{eq:density-mixture}
    f(\cdot) = \sum_{k \in \mathcal{K}} p_k f_{k}(\cdot),
\end{equation}
where $p_k$ is the probability of points belonging to type $k$, and $f_k(\cdot)$ is the spatial density function for type $k$. In the context of our application, $\mathcal{S}$ is the continuous age of individuals under study, $\mathcal{S}=[15,50)$. Each point $\mathbf{s}_i = (a_{i1}, a_{i2})^T$ corresponds to the ages of the two individuals forming a pair, ordered by gender and the latent types are $\mathcal{K} = \{ -1, 0, 1\}$, corresponding to female-to-male transmission, no transmission and male-to-female transmission.  For instance $p_{1}$ corresponds to the proportion of male-to-female transmission events among all pairs of individuals being considered, and $f_1(\cdot)$ corresponds to the 2D function that captures the across-age transmission pattern with male sources and female recipients.

\subsection{Infinite Gaussian mixture models for the typed intensity functions}

There are various choices to model the structure of the density functions $f_k(\cdot)$. To balance simplicity and flexibility, we choose a Dirichlet process (``DP'') Gaussian mixture model (``DPGMM'') consisting of infinitely many bivariate Gaussian components  $f_k(\cdot)$. Specifically, for each point $\mathbf{s}_i$, if its type label $c_i = k$, then
\begin{equation}
\label{eq: DP-GMM}
    \mathbf{s}_{i} \mid c_i = k \sim N(\theta_{ki},
    \Sigma_{ki}), \quad  (\theta_{ki}, \Sigma_{ki}) \sim G_k, \quad G_k \sim DP(\alpha_k, G_0).
\end{equation}
Here $G_k$ represents the (infinite) mixture of bivariate normal models for type $k$, and $\theta_{ki}$ and $\Sigma_{ki}$ are the mean vector and covariance matrix for the bivariate Gaussian component that $\mathbf{s}_i$ belongs to. In practice, Dirichlet process mixtures are often treated as a finite mixture but with a flexible number of components. Indeed, the above defined model may be expressed equivalently in terms of each density function $f_k(\cdot)$,
\begin{equation}
    \label{eq: DPGMM-GMM-express}
    f_k(\cdot) = \sum_{h=1}^{H_k} w_{kh} 
 \cdot \text{dBVN}(\cdot; \theta_{kh}, \Sigma_{kh}),
\end{equation}
where $H_k$ denotes the number of ``active'' components, or total number of unique components generated by the DP, and each $(\theta_{kh}, \Sigma_{kh})$ is a \emph{unique} Gaussian component for the type-$k$ density. Here, $\text{dBVN}((s_{1},s_{2}); \theta, \Sigma)$ denotes the probability density of a bivariate normal distribution with mean $\theta$ and covariance $\Sigma$. 

\subsection{Model for type-dependent marks}

We next describe how the observed point patterns and the associated marks are connected under the typed point process model. We presume that the marks $\mathbf{x}_i$ associated with each point $\mathbf{s}_i$ should provide information on the true event type $c_i$ of each point. This prompts us to model the distribution of the marks conditional on the event type. In our application the marks $\mathbf{x}_i = (\ell_i, d_i)^T$ are two dimensional vectors with entries taking values in $(0,1)$, and we assume the following type-dependent distribution for $\mathbf{x}_i$ conditional on the type value $c_i = k$ ($k \in \{-1, 0, 1\}$): 
\begin{equation}
\label{eq:model-scores-joint}
    p(\mathbf{x}_i \mid c_i = k) = \phi_{k}( (\ell_i, d_i)^T) = dN(\text{logit}( \ell_i); \tilde\mu_{\ell,i}(k), \sigma_{\ell}^2) 
 \times dN(\text{logit}( d_i ); \tilde\mu_{d,i}(k), \sigma_{d}^2).
\end{equation}
Here, $dN(\cdot; \mu, \sigma^2)$ denotes a univariate normal density function with mean $\mu$ and variance $\sigma^2$, and $\text{logit}(x)$ denotes the logit transformation on $x \in (0,1)$. We further specify type-dependent normal means $\tilde\mu_{\ell,i}(k)$ and $\tilde\mu_{d,i}(k)$:
\begin{align}
    \label{eq:mu-ell}
    \tilde\mu_{\ell,i}(k) &= \mu_{\ell}\mathbbm{1}\left[k \neq 0\right] ,\\
    \label{eq:mu-d}
    \tilde\mu_{d,i}(k) &= \mu_{d} \mathbbm{1}\left[k = 1\right] + \mu_{-d}\mathbbm{1}\left[k = -1\right].
\end{align}
Intuitively, a larger linkage score $\ell_i$ indicates stronger evidence for a transmission event, and a larger direction $d_i$ indicates higher confidence for a male-to-female transmission. Thus, with $\mu_{\ell} > 0$, \eqref{eq:mu-ell} implies that the linkage score $\ell_i$ is likely to exceed $0.5$ for a real transmission event ($c_i \neq 0$); similarly, with $\mu_{-d}<0<\mu_{d}$, \eqref{eq:mu-d} implies that $d_i$ is likely larger than $0.5$ for a male-to-female event ($c_i=1$) but smaller than $0.5$ for a female-to-male event ($c_i=-1$). Note that such design uses the property  $\text{logit}(0.5) = 0$.

\subsection{The complete data likelihood}
From the descriptions in the previous section, we see that the two key components in the model---the spatial process and the marks distribution ---are linked through the latent types $c_i$. Conditional on the latent type $c_i = k$, a point $i$ contributes the term $\gamma p_k f_k(\mathbf{s}_i) \times \phi_k(\mathbf{x}_i)$ to the data likelihood, and therefore the likelihood is merely a product of all such terms for $i = 1,2,\ldots, N$. Thus, we first construct the likelihood function given the ``complete'' data, which include the coordinates of all $N$ points in the set $\mathbb{S}$, the observed signals $\mathbf{x}_i$ \emph{as well as the type $c_i$} for all $i = 1, \ldots, N$. We have essentially derived all components in the exposition above, comprising the expression
\begin{align}
    \label{eq: general-likelihood}
    L(\Theta; \{\mathbf{x}_i\}, \{c_{i}\}, \{\mathbf{s}_{i}\}) &= \gamma^N \frac{e^{-\gamma}}{N!}\prod_{k \in \mathcal{K}} \prod_{i: c_i = k} p_k f_k(\mathbf{s}_{i}) \phi_k(\mathbf{x}_i)\\
    \label{eq:likelihood-DPGMM}
    =& \prod_{i=1:N} dN(\text{logit}( \ell_i); \tilde\mu_{\ell,i}(c_i), \sigma_{\ell}^2) 
dN(\text{logit}( d_i ); \tilde\mu_{d,i}(c_i), \sigma_{d}^2) \\
    &\times \gamma^N \frac{e^{-\gamma}}{N!} \prod_{k\in\mathcal{K}} \prod_{i: c_{i}=k} \left(p_k \sum_{h=1}^{H_k} w_{kh} \text{BVN}((s_{i1},s_{i2}); \theta_h, \Sigma_h)\right). \nonumber
\end{align}
Here, the model parameters are $\Theta = \{\gamma, \mathbf{p}, \boldsymbol\mu, \sigma_{\ell}^2, \sigma_{d}^2, \{(\theta_{kh}, \Sigma_{kh})\}, \{\alpha_k\}\}$, where we further abbreviate $\mathbf{p} = (p_{-1}, p_0, p_1)^T$, and $\boldsymbol\mu = (\mu_{\ell}, \mu_{d}, \mu_{-d})^T$).

\section{Bayesian inference with data augmentation}
\label{sec: inference}

We employ an efficient data-augmented Bayesian inference scheme to learn the unknown parameters $\Theta$ in the proposed typed point process model from observed data. Inference would be straightforward if all aspects of the model were observable, giving access to the complete data likelihood specified in \eqref{eq:likelihood-DPGMM}. That is, if the $c_i$'s were known, the terms corresponding to the spatial point process and marks  completely factorize in the likelihood function, as shown in \eqref{eq:likelihood-DPGMM}. Parameter inference would reduce to standard procedures for Dirichlet Process Gaussian mixture models~\citep{rasmussen1999infinite}. 

However, the types $c_i$ for each data point $i$ are not observed in our data setting. Because it is not clear whether direct marginalization is possible, we propose a data augmentation inference scheme that expands the target posterior to include the unobserved $c_i$'s as unknowns. Conditioning on a given setting of their values then gives us access to the tractable complete data likelihood function. That is, our algorithm samples from the joint posterior density of the model parameters $\Theta$ together with the unobserved types $c_i$'s: 
\begin{align}
    p(\Theta, \{c_i\} \mid \{\mathbf{x}_i\},  \{\mathbf{s}_{i}\}) &\propto L(\Theta; \{\mathbf{x}_i\}, \{c_{i}\}, \{\mathbf{s}_{i}\}) p_0(\Theta).
\end{align}
Here $p_0(\Theta)$ denotes the joint prior distribution for parameters $\Theta$, which we will detail in the sequel. The data-augmented inference framework is employed through a Bayesian Markov chain Monte Carlo (MCMC) sampler, which can be roughly divided into two major components in each iteration: (1) sample or update parameters $\Theta$ conditioned on configurations of the $\{c_i\}$'s from the conditional posterior distribution $p(\Theta \mid \{\mathbf{x}_i\}, \{c_{i}\}, \{\mathbf{s}_{i}\})$; and (2) sample $c_i$ for each $i$ given values of $\Theta$ from $p(c_i \mid \Theta, \mathbf{x}_i, \mathbf{s}_{i})$, utilizing the complete data likelihood in \eqref{eq:likelihood-DPGMM}. To improve efficiency of the MCMC sampler, we prescribe conjugate or semi-conjugate priors whenever possible, enabling straightforward Gibbs sampling by exploiting full conditional posterior densities available for almost all parameters. Below, we detail these prior choices and discuss each step of the MCMC sampler. A pseudocode summary of our sampler appears as Algorithm 1 in Web Appendix A.

\subsection{Scale parameter $\gamma$} 
Due to the scale decomposition in \eqref{eq:decompose-all}, sampling the parameter $\gamma$ is straightforward and can in fact be done independently of the Markov chain that samples the remaining parameters. That is, if we assign a Gamma prior $\gamma\sim Ga(a_0, b_0)$, we may directly draw samples using \[ \gamma \mid {\{\mathbf{x_i}\}, \{\mathbf{s_i}\}} \sim Ga(\alpha_0+N, \beta_0 + 1) . \]

\subsection{Mark distribution parameters $\boldsymbol{\mu}$, $\sigma_{\ell}^2$, and $\sigma_{d}^2$} 
Conditioned on type labels $\{c_i\}$, the signal distributions for $\ell_i$'s and $d_i$'s \eqref{eq:model-scores-joint} are simple univariate normal models. 
Assuming diffuse priors $\mu_{\ell}, \mu_d \sim \text{Unif}((0,\infty))$, $\mu_{-d} \sim \text{Unif}((-\infty,0))$ and inverse-Gamma priors $\, \sigma_{\ell}^2,\sigma_{d}^2 \sim \text{inv-Gamma}(\nu_0/2, \nu_0\sigma_0^2/2)$ \,, parameter updates only require straightforward Gibbs steps from the full conditionals. 
For example, for the linkage score parameters $\mu_{\ell}$ and $\sigma^2_{\ell}$, we draw
\begin{align*}
    \mu_{\ell} \mid \sigma^2_{\ell}, \{\ell_i\}, \{c_i\} &\sim N_{(0,\infty)}\left(\sum_{i:c_i\neq 0} \text{logit}(\ell_i), \sigma^2_{\ell}/N_+\right)\\
    \sigma^2_{\ell} \mid \mu_{\ell}, \{\ell_i\}, \{c_i\} &\sim \text{inv-Gamma}\left(\frac{\nu_0+N_+}{2}, \frac{\nu_0\sigma_0^2+\sum_{i:c_i\neq 0} (\text{logit}(\ell_i)-\mu_{\ell})^2}{2}\right).
\end{align*}
Here $N_+ = \sum_{i=1}^N \mathbbm{1}(c_i \neq 0)$ is the total number of real transmissions given the $\{c_i\}$ configurations, and $N_{(0,\infty)}$ denotes a normal distribution truncated on the positive real line. For the direction score parameters $\mu_{d}, \mu_{-d}$ and $\sigma_d^2$, sampling steps are almost completely analogous.

\subsection{Type probabilities $\mathbf{p}$}
Assuming a Dirichlet prior for the probability vector $\mathbf{p} = (p_{-1},p_{0},p_1)$, $\mathbf{p} \sim \text{Dir}(q_{-1},q_{0},q_1)^T$ and given configurations for $\{c_i\}$, we can draw
\begin{equation*}
    \mathbf{p} \mid \{c_i\} \sim Dir\left( q_{-1} + N_{-1}, q_{0} + N_0,q_1+N_1\right),
\end{equation*}
where $N_{k} = \sum_{i=1}^N \mathbbm{1}(c_i = k)$ is the total number of data points belonging to type $k$. 

\subsection{DP precision parameter $\alpha_k$ and component weights $w_{kh}$}
For the precision parameter $\alpha_k$ and weights $w_{kh}$ for each type $k$, we adopt a truncated DP mixture model with a large maximum number of mixtures to approximate the infinite DP mixture, a technique described in Section 3.2 and in the Appendix of \cite{ji2009spatial}. More specifically, we use an auxiliary sampling trick introduced in \cite{escobar1995bayesian} to update the precision parameter $\alpha_k$'s for $k=0,-1,1$). By introducing an additional auxiliary parameter to sample along with each $\alpha_k$, the conditional posterior distribution for $\alpha_k$ can be reduced to a mixture of two Gamma distributions, which conveniently transforms its sampling step to a simple Gibbs step. For complete technical details, we refer the reader to  \cite{ji2009spatial} and \cite{escobar1995bayesian}. 

\subsection{BVN mixture components $(\theta_{kh}, \Sigma_{kh})$}
For the bivariate normal mixture model of each type $k$, we introduce a component latent indicator $z_i$ for each data point $i$ (that belongs to type $k$) such that $z_i=h$ indicates point $i$ belongs to component $h$. Assuming semi-conjugate priors $\theta_{kh} \sim \text{BVN}(\theta_0, \Sigma_0)$ and $\Sigma_{kh} \sim \text{inv-Wishart}(\nu, S_0)$, we then iteratively update the $z_i$'s and $(\theta_{kh}, \Sigma_{kh})$'s using
\begin{align*}
    Pr(z_i = h \mid w_{kh},\theta_{kh}, \Sigma_{kh}, \mathbf{s_i}) &\propto w_{kh} \cdot \text{dBVN}(\mathbf{s}_{i} \mid \theta_{kh}, \Sigma_{kh})\\
    \theta_{kh} \mid \Sigma_{kh}, \{z_i\}, \{\mathbf{s_i}\} &\sim \text{BVN}\bigg((m_h\Sigma_{kh}^{-1}+\Sigma_0^{-1})^{-1}\left(\sum_{i:z_i=h}\Sigma_{kh}^{-1}\mathbf{s_i} + \theta_0\Sigma_0^{-1}\theta_0\right),\\
    &\quad\quad\quad\quad\quad (m_h\Sigma_{kh}^{-1}+\Sigma_0^{-1})^{-1}\bigg);\\
    \Sigma_{kh} \mid \theta_{kh},\{z_i\},\{\mathbf{s_i}\} &\sim \text{inv-Wishart}\left(\nu+m_h, \left(S_0^{-1}+\sum_{i:z_i=h}(\mathbf{s_i}-\theta_{kh})(\mathbf{s_i}-\theta_{kh})^T\right)^{-1}\right).
\end{align*}
Here $\text{dBVN}(\cdot \mid \theta, \Sigma)$ is the density function of a bivariate normal with mean $\theta$ and covariance matrix $\Sigma$, and $m_h = \sum_{i=1}^N \mathbbm{1}(z_i = h)$ is the total number of data points belonging to spatial component $h$. 

\subsection{Type indicators $c_i$}
The type $c_i$ can be sampled for each data point $i$ conditioned on all other parameter values and the observed data via its full conditional distribution
$Pr(c_i = k \mid \Theta, \mathbf{s}_{i}, \mathbf{x}_i) \propto p_kf_k(\mathbf{s}_{i}) \phi_k(\mathbf{x}_i)$.

\section{Simulation studies}
\label{sec:simulation-experiments}

In this section, we validate the modelling and inference framework for marked point patterns through synthetic experiments. With our application in mind, we focus on assessing whether the typed point process model can accurately recover simulated patterns of HIV transmission flows between men and women across different ages. We further explore the ability of the model to infer these flows at increasingly smaller sample sizes. Throughout this section, we focus on two contemporary questions of particular epidemiological interest, and investigate the ability of the typed point process model in recovering and differentiating between two competing scenarios. 

First, one general finding of recent HIV transmission flow studies is that there tend to be more infections that originate from men than from women~\citep{hall2021demographic,bbosa2020phylogenetic,ratmann2020quantifying}. This prompts us to explore whether the typed point process model can accurately recover parameters and distinguish between a scenario in which men drive 50\% of infections (``MF-50-50") and a scenario in which men drive 60\% of infections (``MF-60-40''). We explore $5$ different sample sizes (i.e., numbers of likely transmission pairs) with $N = 100, 200, 400, 600$ and $800$. For each scenario and each sample size, $100$ independent data sets were generated, and the typed point process model fitted to each. Details on the simulations and scenarios are fully described in Web Appendix B. 

Second, there is broad interest in the age distribution of the male sources of HIV infections, especially for adolescent girls and young women aged 15 to 24, due to the high incidence rates in these women~\citep{risher2021age}. This prompts us to investigate if the model can distinguish between two scenarios in which either younger men of approximately 25 years (within $\pm 5$ years age difference of adolescent and young women) contribute 60\% to infections in women aged 15-24 years and older men of approximately 35 years 30\%, and other men 10\% (``SAME AGE"); versus the alternative scenario that younger men of approximately 25 years contribute 30\%, and older men of approximately 35 years 60\%, and other men 10\% (``DISCORDANT AGE"). We varied the sample sizes, generated simulations and fitted the model as before, with complete details described in Web Appendix B.

\begin{figure}
    \centering
    \includegraphics[width = \textwidth]{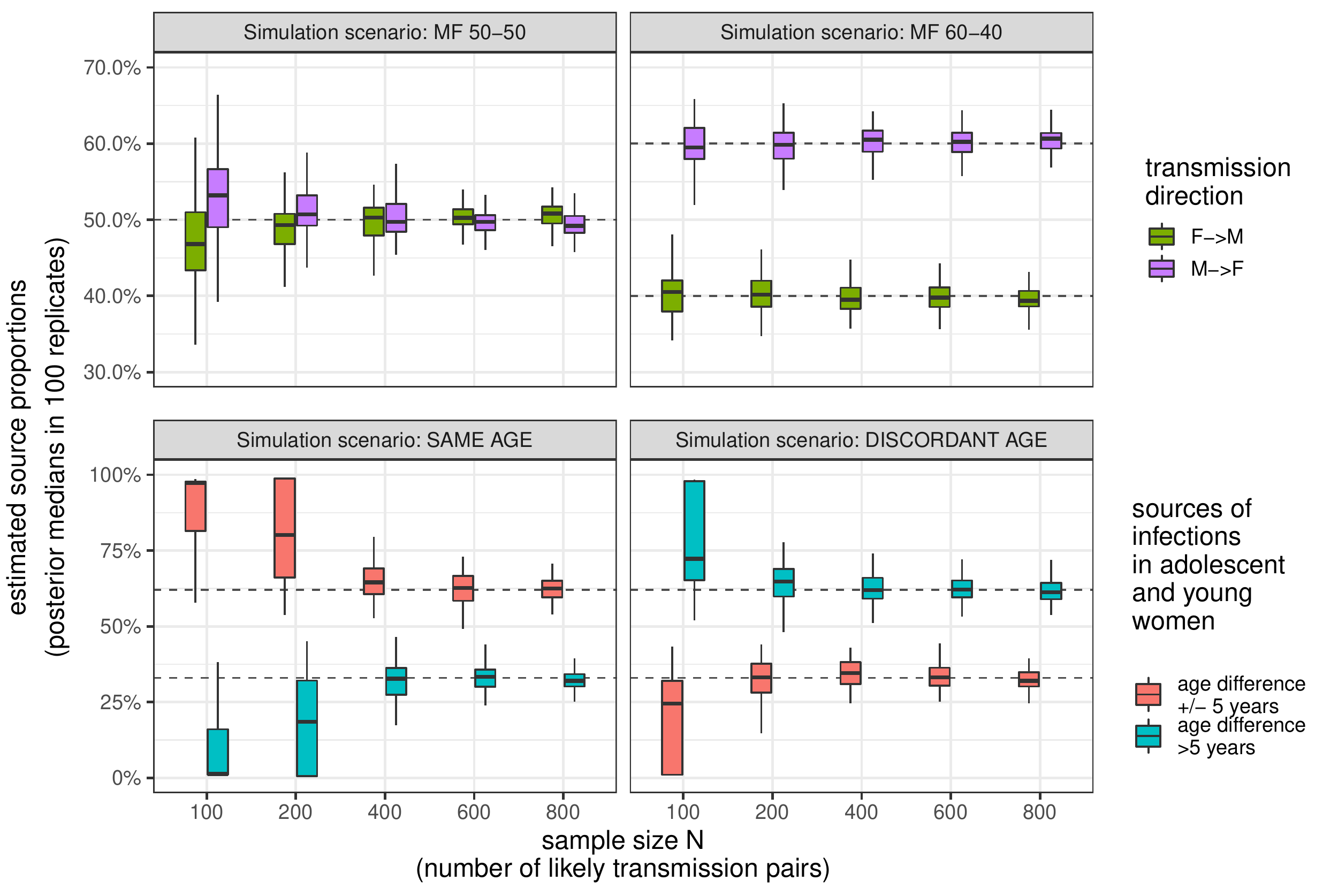}
    \caption{{\bf Performance of the typed point process model in recovering simulated transmission flow patterns.} Inference using the typed point process model was performed on each of the 100 simulated data sets for each scenario. {\bf A} Boxplot of the posterior mean estimate of transmissions from men in 100 replicate simulations for the MF-50-50 (left panel) and MF-60-40 scenarios (right panel). Throughout, the dashed lines mark the true values that underpin the simulated data. The x-axis shows the sample size of simulated data points, which represent the number of phylogenetically closely related pairs of individuals identified through phylogenetic deep-sequence analyses. {\bf B} Boxplot of the posterior mean estimate of transmissions from men of similar age (shown in red) and older age (shown in blue) to infection in adolescent and young women aged 15-24 in 100 replicate simulations for the ``SAME AGE'' and ``DISCORDANT AGE'' scenarios. As before the dashed lines mark the true values that underpin the simulated data and the x-axis shows results for different sample sizes.}
    \label{fig:weights_props_by_N_joint}
\end{figure}

Figure~\ref{fig:weights_props_by_N_joint} (top panel) summarizes our findings on the male-female simulation experiments (MF-50-50 and MF-60-40). Our findings reveal that the model is able to successfully distinguish between the two competing epidemiological scenarios with sample size $N \geq  200$, and produces estimates that are accurate within a $\pm 5$\% error margin for sample sizes $N \geq 200$. Figure~\ref{fig:weights_props_by_N_joint} (bottom panel) illustrates our findings on the age-specific sources experiments (SAME-AGE, DISCORDANT-AGE). This is a substantially more difficult inference problem because the target quantities relate to a smaller subgroup of the entire source population. With a small sample ($N$ in the range $100-200$), the relative relationship between the two proportions is inferred correctly in all experiments. However, the actual quantitative estimates and differences can be far from the truth, and tend to be over-estimated. For sample sizes of $N\geq 400$, the quantitative estimates become accurate for practical purposes. The observed over-estimation is likely due to the parsimony induced by the Dirichlet Process priors that are known to prefer assigning data points to the largest existing clusters when there are not enough data to admit a new mixture model component. As a result, more points tend to be attributed to the component with the highest weight when $N$ is small; this effect is mitigated as $N$ increases. Additional analyses on simulated experiments are included in Web Appendix B. These include numerical convergence and mixing analyses, which are generally quite satisfactory, as well as Bayesian coverage analyses, which indicate that the posterior distributions of the target quantities are well calibrated.

\section{Case study}
\label{sec: case-study}
We are now ready to apply the typed point process model to study demographic and population-based HIV deep-sequence data collected from the Rakai Community Cohort Study in Southeastern Uganda between August 2011 and January 2015~\citep{ratmann2019inferring, ratmann2020quantifying, xi2021inferring}. Our aim is to reconstruct transmission flows by gender and continuous age between 15 and 50 years. This case study is challenging in the sense that phylodynamic analyses using standard HIV consensus sequences have difficulty inferring flow patterns for more than a handful of age groups: many contemporary analyses of age-specific transmission flows remain limited to a resolution of 5-year or 10-year age bands~\citep{Scire2020Improved,bbosa2020phylogenetic,le2019hiv,de2017transmission, ratmann2020quantifying}.

We revisit the HIV deep-sequence dataset considered by  ~\cite{xi2021inferring}, and compare our approach applied to the complete dataset without appealing to heuristic pre-classification that results in utilizing only a small subset of data. More specifically, we will use the typed point process model to analyze all pairs of closely phylogenetically related individuals introduced in Section~\ref{sec: data-model}. After filtering out only pairs of individuals with a linkage score $< 0.2$ (i.e., pairs with very weak evidence), we retain a dataset of 526 pairs that represent potential transmission events. For comparison purposes, we also repeat the analysis under the heuristic thresholding approach following~\cite{xi2021inferring}; this pre-processing step retains only retains a subset of those pairs of individuals as ``high-confidence'' transmission pairs. Referring to the first as the ``full analysis with latent event types'' (or ``full analysis'' in short), we fit the typed point process model to all data, simultaneously inferring the three latent event types in our application, $k=0$ for no transmission event, $k=1$ representing male-to-female transmission, and $k=-1$ for female-to-male transmission (recall the Methods section). In the second ``subset analysis with fixed event types'' (``subset analysis'' in short), we retain similarly as in~\citep{xi2021inferring, hall2021demographic} only those data with linkage scores $>0.6$ and a direction score $>0.5$  as a male-to-female transmission pair, and those with linkage scores $>0.6$ and a direction score $\leq 0.5$ as a female-to-male transmission pair. In the full analysis, all 526 data points will be used, while the subset analysis considers only 367 pairs. This is a larger subset of data retained compared to~\cite{xi2021inferring} which consists of 238 pairs of individuals, where only ``high-confidence'' potential transmission pairs with direction score $<0.33$ or $>0.67$ were retained for analysis. In both analyses, we adopt the same priors as described in Section~\ref{sec:simulation-experiments} and run the MCMC algorithm detailed in Section~\ref{sec: inference} for 3000 iterations after 1000 burn-in steps.

\subsection{Learning the latent event types is computationally feasible}

We are able to infer both the transmission flow surfaces of interest, $f_1$ and $f_{-1}$ in \eqref{eq:density-mixture}, and the unknown event types $\{c_i; i= 1,\dotsc, N\}$ for each observed data point without issues in numerical convergence or mixing. On a laptop with a 4-core Intel CPU, the full analysis with the typed point process model including the latent event types takes less than 10 minutes, which is a considerable improvement in computational efficiency compared to using the semi-parametric Poisson count model on 1-year age discretized bands~\cite{xi2021inferring}, where 4000 total iterations requires about 30 hours. This speed-up is expected, as our typed point process model greatly simplifies computation through a continuous spatial point pattern specification that only requires likelihood evaluations on the observed $526$ data points. In comparison, the Poisson count model in \cite{xi2021inferring} involves likelihood evaluations on all $2\times (50-15)^2=2450$ grid cells, many of which contain structural zeros.

\subsection{Learning latent event types results in more data used for inference of transmission flows}

In comparison to the subset analysis, the full analysis appears to attribute more pairs of individuals to type $1$ (male-to-female transmission) and to type $-1$ (female-to-male transmission). We present the posterior probability $p_k$ for each type as inferred by the full analysis and by the subset analysis in Table~\ref{tab: type-proportions}. Multiplying the estimated $p_k$'s by sample size $526$, we also calculate the number of pairs, $N_k$, that each analysis has attributed to each type. Notably, the full analysis learns a $p_0$ (proportion of ``no transmission'') that is significantly lower than that under the fixed event type subset analysis. This indicates that, by probabilistically inferring event types instead of heuristically pre-classifying them, our proposed approach is utilizing more phylogenetically close pairs to learn the transmission flows. In short, our approach not only reflects uncertainty in event types, but also makes more efficient use of phylogenetic and demographic evidence in data.

\begin{table}[ht]
\caption{{\bf Proportions and numbers of inferred event types.} Posterior mean estimates with 95\% credible intervals.}
\centering
\begin{tabular}{l|cc|cr}
  \hline
\multirow{ 3}{*}{Type} & \multicolumn{2}{l|}{\textbf{Full analysis}} & \multicolumn{2}{l|}{\textbf{Subset analysis}}\\
& \multicolumn{2}{l|}{\textbf{with latent event types}} & \multicolumn{2}{l|}{\textbf{with fixed event types}}\\
\cline{2-3} \cline{4-5}
 & $p_k$ & $N_k$ &  $p_k$ & $N_k$ \\ 
  \hline
Male-to-Female & 46.3\% (39.4\%, 53.1\%) & 244 (207, 279)  & 35.7\% & 188 \\ 
transmission $(k=1)$& & & & \\
Female-to-male & 35.0\% (28.5\%, 42.0\%) & 184 (150, 221) & 29.5\% & 155 \\ 
transmission $(k=-1)$ & & & & \\
No transmission & 18.6\% (9.4\%, 29.2\%) & 98 (49, 154) & 34.8\% & 183 \\ 
$(k=0)$ & & & &\\ 
   \hline
\end{tabular}
\label{tab: type-proportions}
\end{table}

\subsection{Inferred age of male and female sources of HIV transmission.}

\begin{figure}
    \begin{subfigure}{0.98\textwidth}
    \includegraphics[width=0.48\textwidth]{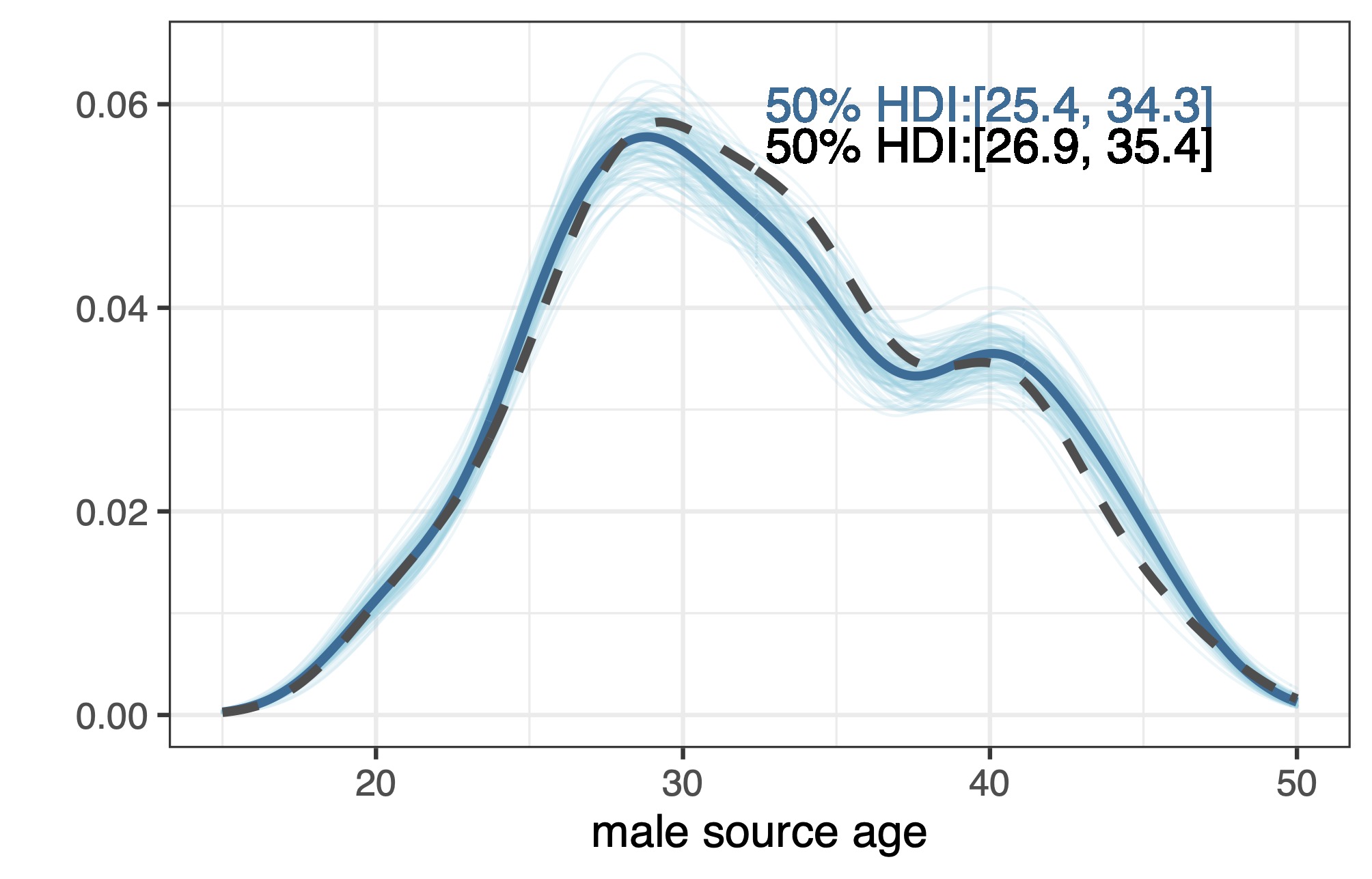}
    \includegraphics[width=0.48\textwidth]{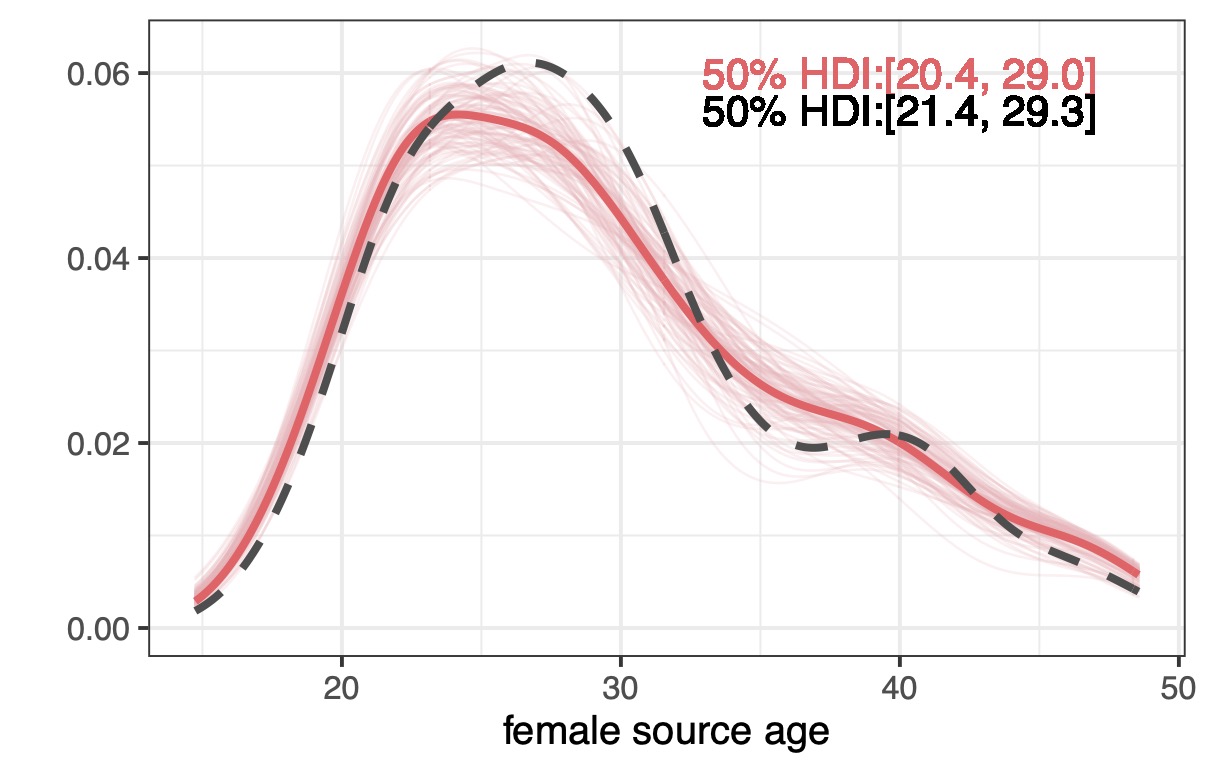}
    \caption{{\bf Age distributions of male and female sources of HIV infections in Rakai, Uganda during the 2011-2015 observation period.}
    The left panel shows the estimated age of the male sources and the right panel the estimated age of the female sources.}
    \label{fig:age-distributions-general}
    \end{subfigure}
    ~
    \vspace{0.15in}
    
    \begin{subfigure}{0.98\textwidth}
        \centering
    \includegraphics[width=0.48\textwidth]{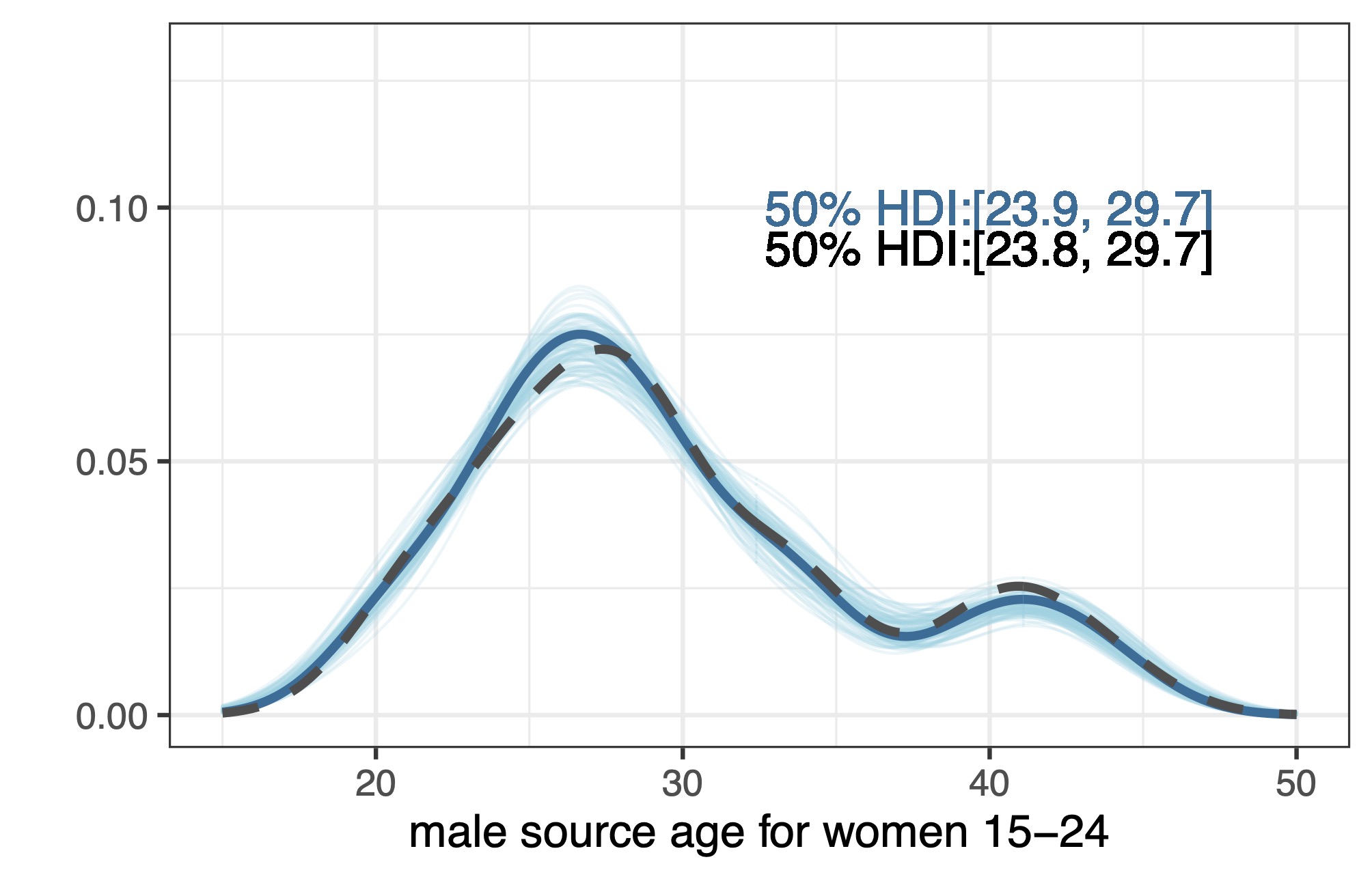}
    \includegraphics[width=0.48\textwidth]{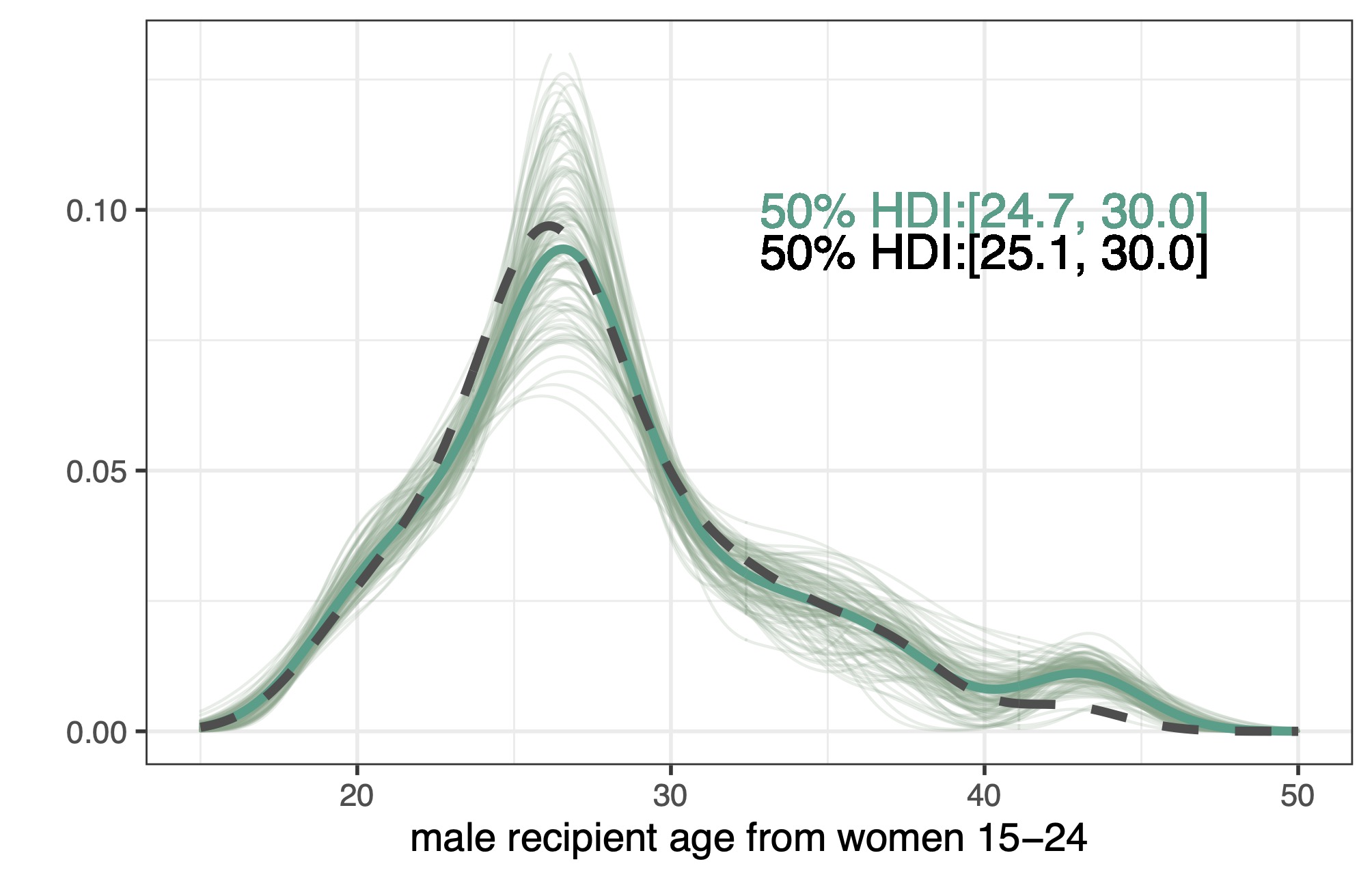}
    \caption{{\bf Age distributions of male sources and recipients of HIV infections in women aged 15-24.}
    Left panel shows the age distribution of \emph{male sources}, and right panel characterizes age distribution of \emph{male recipients}, for women aged 15-24.}
    \label{fig:age-distributions-young-women}
    \end{subfigure}
    
    \caption{\textbf{Age distributions of sources and recipients, for the general population (subplot \ref{fig:age-distributions-general}) and for women aged 15-24 (subplot \ref{fig:age-distributions-young-women})}. In each panel, the colored lines represent density curves of the age of sources/recipients for 100 posterior samples from the inferred, smooth transmission flow intensity surface of the typed point process model in the full analysis with latent event types. The thicker curve indicates the posterior mean density curve. 
    The black dashed curve illustrates the posterior mean density curve in the subset analysis with fixed event types. 
    50\% highest density intervals (HDIs) are marked in text, with colored text indicating the HDIs inferred in the full analysis and black text indicating the HDIs inferred in the subset analysis.}
\end{figure}

With the typed point process model, we are able to estimate the sources of male and female HIV infection events in terms of the continuous age of the sources (Figure~\ref{fig:age-distributions-general}). This is an important advantage over other approaches through which the age of the sources can be characterised by 5-year or 10-year age bands~\citep{Scire2020Improved,bbosa2020phylogenetic,le2019hiv,de2017transmission}, because at such coarse scale discretization introduces artifacts that obfuscate differences in transmission modes that are important to differentiate from a public health perspective~\citep{xi2021inferring}. Figure~\ref{fig:age-distributions-general} addresses the following question about HIV transmission --- which age groups contribute  most to HIV transmission at the population-level? The colored solid curves and text illustrate the inferred age distribution of the male and female sources in the full analysis with latent event types, while the dark dashed lines and dark text summarize results for the subset analysis with fixed event types. We find that the inferred age distribution of the male and female sources are very similar in both analyses. 

More specifically, during the observation period 2011-2015, male sources tended to be consistently older than female sources. The estimated 50\% highest posterior density intervals (HDIs) learned in the full analysis with latent event types are ages $[25.4, 34.3]$ for male sources and ages $[20.4, 29.0]$ for female sources. In comparison, the 50\% posterior HDI inferred in the subset analysis with fixed event types are $[26.9, 35.4]$ for the age distribution of male sources, and  $[21.4, 29.3]$ for female sources. This indicates that estimation uncertainty in quantities of key epidemiological interest did not decrease in the full analysis. Rather, we find that the full analysis using latent event types better reflects the actual uncertainty in key quantities, as the typed point process model explicitly accounts for uncertainty in the event types that underpin each data point.

Both the full and the subset analyses indicate a characteristic peak in transmission from men around 30 years of age, and a long, pronounced tail of transmissions from older men beyond age 40. In this regard, we find that the age distribution of the female sources differs qualitatively, as the probability of transmissions from women declines more rapidly with age, reaching similar levels by age 30 in women compared to age 45 in men. In the full analysis with latent event types, these observations on the age of the female sources are more strongly supported, as in the full analysis the entire age distribution of the female sources is slightly shifted towards younger ages compared to the age distribution of the female sources in the subset analysis with fixed event types.

\begin{figure}

    \begin{subfigure}{0.98\textwidth}
    \centering
    \includegraphics[width=0.48\textwidth, page=3]{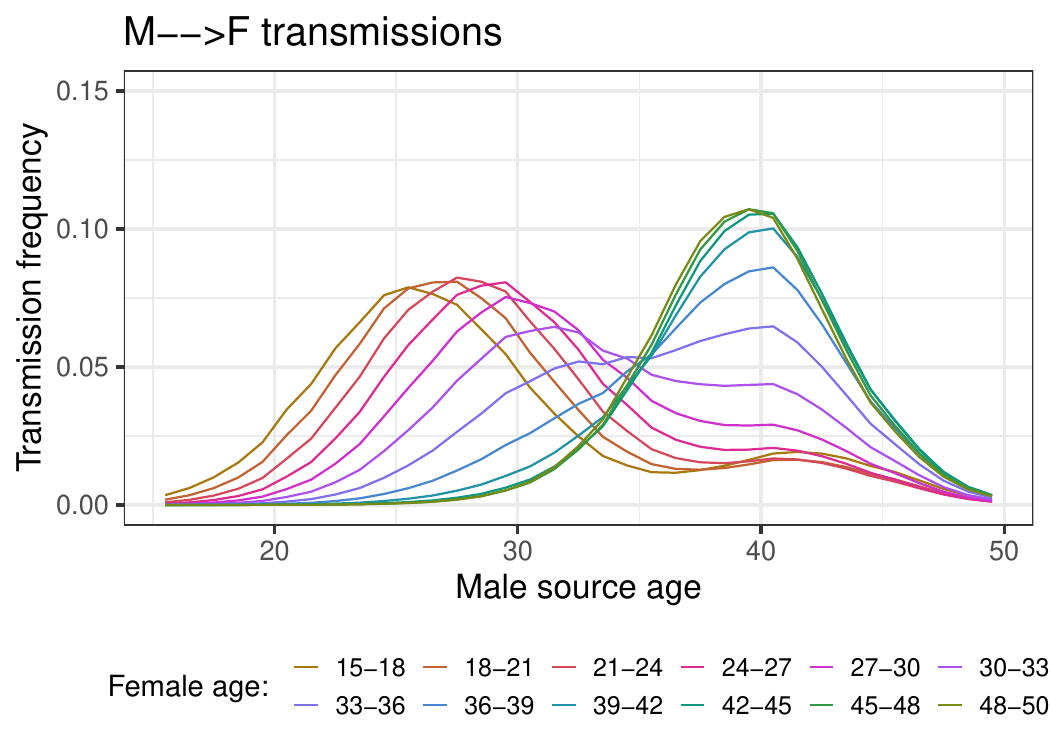}
    \includegraphics[width=0.48\textwidth, page=4]{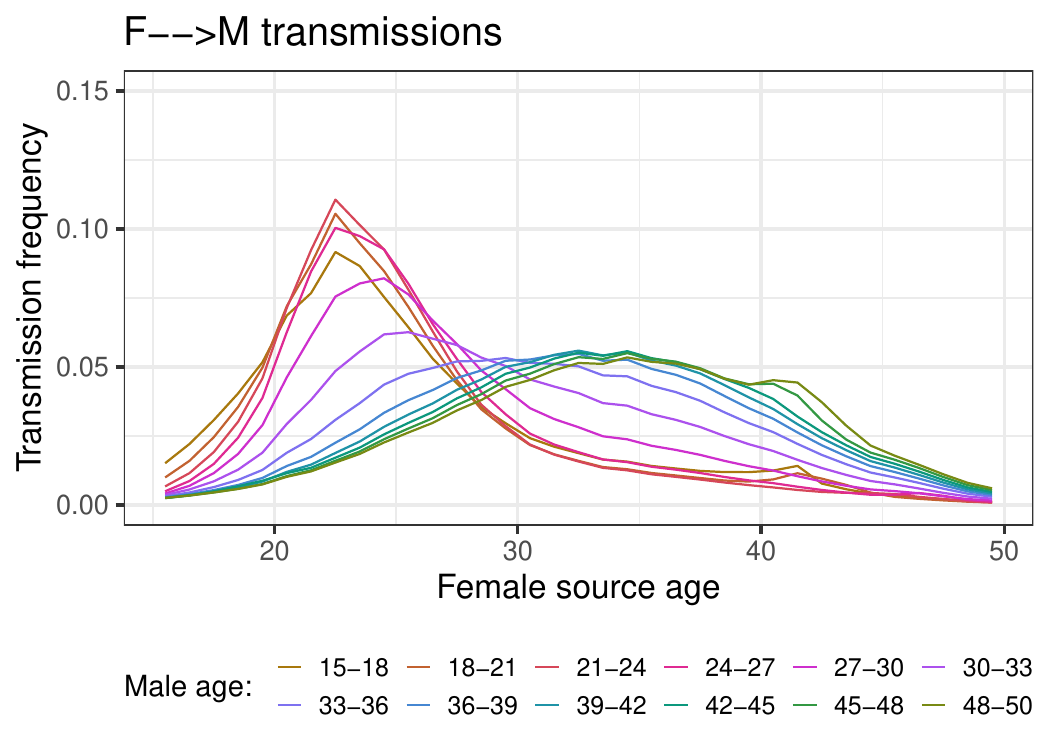}

    \caption{\textbf{Age distributions of sources for recipients in different 3-year age groups.} Each curve represents the learned relative frequencies of sources responsible for transmissions within each recipient age group. }
    \label{fig: age-source-distribution-3year}
    \end{subfigure}
    
    ~

    \vspace{0.15in}

    \begin{subfigure}{0.98\textwidth}
    \centering
    \includegraphics[width=0.48\textwidth, page=3]{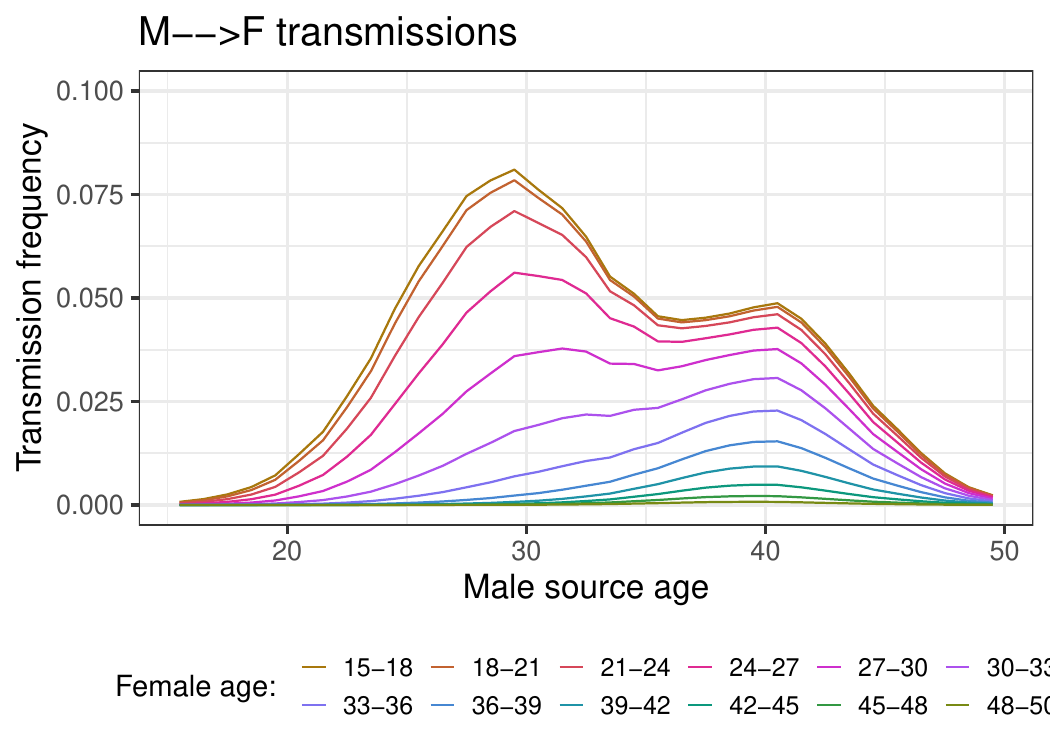}
    \includegraphics[width=0.48\textwidth, page=4]{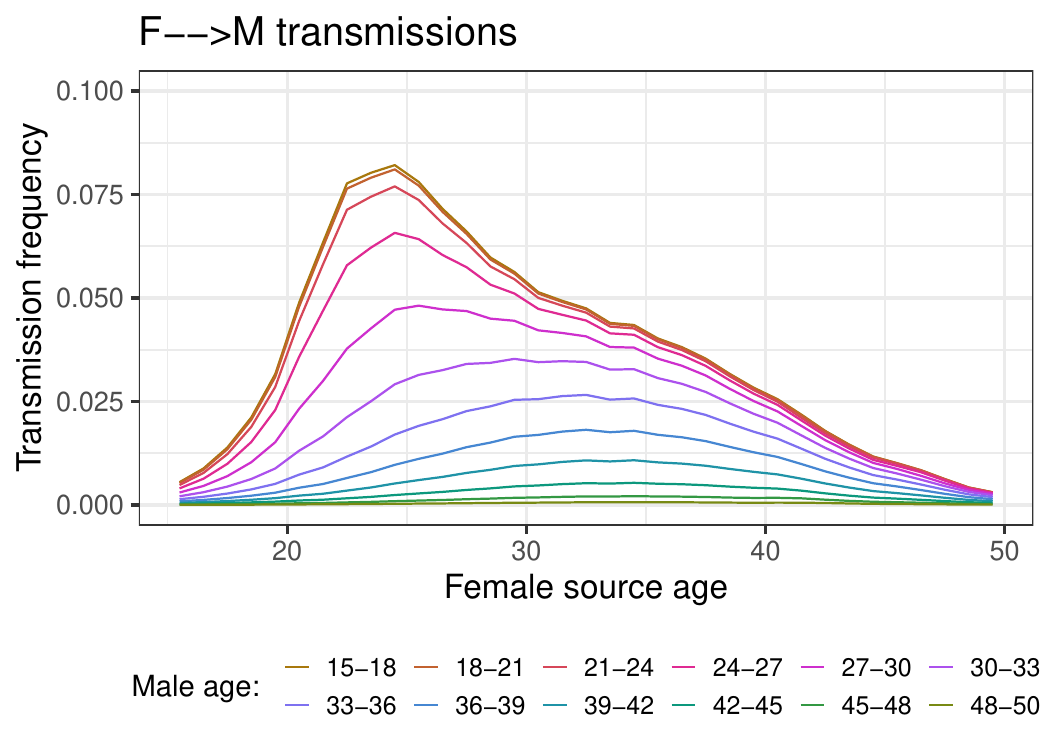}

    \caption{\textbf{Marginal age distributions of sources.} Shown as ``stacked'' curves of age source distributions for each recipient age group (in 3-year age bands). }
    \label{fig: age-source-distribution-stacked}
        
    \end{subfigure}
    
    \caption{\textbf{Age distributions of sources for heterosexual recipients in different age groups (by 3-year age bands).} 
    \textbf{In subplot \ref{fig: age-source-distribution-3year}}, the curves for each recipient age group represent posterior means of the source age distributions learned by the full analysis with latent event types. 
    \textbf{In subplot \ref{fig: age-source-distribution-stacked}}, the ``marginal'' age distributions are shown by stacking up the frequencies of sources for each recipient age group. 
    \textbf{Left} column shows age distributions for male sources in male-to-female transmissions for different female recipient age groups. 
    \textbf{Right} column shows age distributions for female sources in female-to-male transmissions for different male recipient age groups. }
    \label{fig:source-prob-by-age-full}
\end{figure}

We can gain further insight into age-specific transmission dynamics by considering the age profile of the transmitting partners for recipients of a particular age. In  Figure~\ref{fig:source-prob-by-age-full}, the age of a recipient is shown in colour, and the wave on the $y$-axis shows the posterior median contribution of transmissions to recipients of the age indicated in colour. Figure~\ref{fig: age-source-distribution-3year} illustrates that the age profile of sources has a characteristic shape for each recipient age group, and in particular are not simply shifted versions of one age profile. Figure~\ref{fig: age-source-distribution-stacked} demonstrates how the superposition of the age-structured transmission dynamics results in the overall source profile that marginalises out the age of the recipients.

\subsection{Transmissions to and from adolescent and young women}
Next, we focus our attention on adolescent and young women of age 15 to 24 years. 
Specifically, we wish to understand the age distribution of male sources in their infectious, and in turn, that of male recipients for whom these women are the sources of infection. Our rationale is that understanding the age distribution of the male sources and recipients in adolescent and young women can provide further evidence to HIV programs that aim to reduce infections in this age group~\citep{glynn2001young,pettifor2008keep,karim2010preventing,jewkes2010intimate}. One prominent program is the ``Determined, Resilient, Empowered, AIDS-free, Mentored, and Safe'' (DREAMS) program~\citep{saul2018determined} within the US President’s Emergency Plan for AIDS Relief (PEPFAR)~\cite{oliver2012us}. 


Figure~\ref{fig:age-distributions-young-women} illustrates the inferred age distributions of the male sources of infections in adolescent and young women (left), and of the male recipients of transmissions from these women (right) in the full analysis with latent event types (colored curves) and the subset analysis with fixed event types (black curves). While the majority of male sources are men between 24 to 30 years, there is also a notable subgroup of male sources older than 35. In contrast, the male recipients of transmission from adolescent and young women lie more strongly in men aged 24 to 30 years, with a much smaller subgroup of male recipients older than 35. That is, a transmission pathway strongly supported by the 2011-2015 data may begin with transmission from men approximately 5-8 years older than  adolescent and young women, and who then in turn transmit the virus in smaller numbers to men who are again 5-8 years older than themselves. In addition, a second transmission pathway may begin with transmission from men 10-15 years older than adolescent and young women, and who then in turn transmit the virus in smaller numbers to men who are more likely to be 5-8 years older than themselves, rather than 10-15 years older. Our findings for the 2011-2015 observation period thus further corroborate the importance of major HIV prevention programs across Sub-Saharan Africa that have been directing educational and healthcare resources to adolescent and young women, who are especially vulnerable to HIV from a broad age spectrum of male sources~\citep{saul2018determined,UNAIDS2018}.

\begin{figure}
    \centering
    \includegraphics[width=0.98\textwidth]{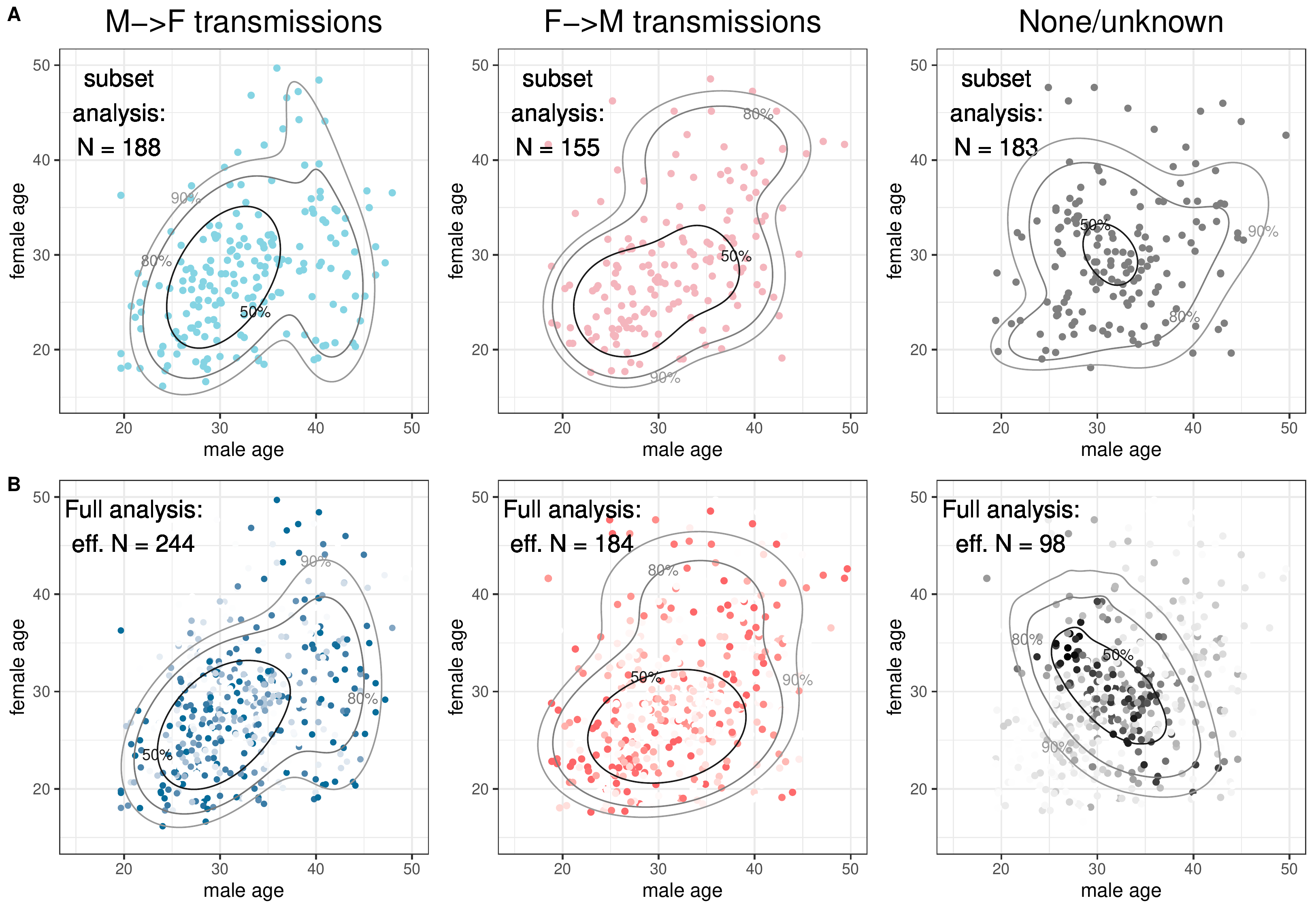}
    \caption{{\bf Comparison of the inferred age structure in transmission flows in the full with latent event types versus the subset analysis with fixed event types.} ({\bf A}) Results in the subset analysis with fixed event types. Source-recipient pairs that were pre-classified by event type (dots) are shown along the posterior median estimate of 50\%, 80\% and 90\% highest probability regions of transmission flows (contours). The number of data points attributed to each type is indicated in the top left corner.  
    ({\bf B}) Results in the full analysis with latent event types. Source-recipient pairs (dots) are shown by posterior event type probabilities (colour intensity) along the posterior median estimate of 50\%, 80\% and 90\% highest probability regions of transmission flows (contours). The ``effective'' number of data points attributed to each type (posterior mean estimate of $N_k$ as in Table~\ref{tab: type-proportions}) is indicated in the top left corner.}
    \label{fig:points-colored-by-probs-with-contours}
\end{figure}

\subsection{In-depth comparison of the full analysis with latent event types with the subset analysis with fixed event types}
We close by examining the more detailed features of the inferred age- and gender-specific transmission flows in the subset analysis with fixed event types (Figure~\ref{fig:points-colored-by-probs-with-contours}A) compared to the full analysis with latent event types  (Figure~\ref{fig:points-colored-by-probs-with-contours}B). In the subset analysis with fixed event types, only a subset of all data points are utilized to learn the latent age structure associated with each type, and all data points are assumed to carry the same weight (as colored by the same shade in each panel). In the full analysis with latent event types, all data points contribute to the learning of the latent age structure associated with each event type, and the contribution of each data point is weighted by the posterior event type probabilities associated with each point (as illustrated with the colour shades of each point). Overall, we find that the learned 2D surfaces of age-specific male-to-female transmission flows and female-to-male transmission flows are very similar in the full analysis with latent event types compared to the subset analysis (contours in Figure~\ref{fig:points-colored-by-probs-with-contours}. This is already indicated by the findings on the marginal age distribution of the male and female sources of infection in Figure~\ref{fig:age-distributions-general}.

However, the inferred latent surface corresponding to no transmission differs substantively between the full and subset analyses, with noticeably more data points in the corners of the compact space attributed to either male-to-female or female-to-male transmission. Data points corresponding to combinations of women around age 20 and men around age 30 are more strongly attributed to male-to-female transmission in the full analysis, in line with the strong center of mass of the inferred male-to-female transmission surface for these age combinations. Data points involving women around age 45 are more strongly attributed to female-to-male transmission, in line with the wider tails of the inferred female-to-male transmission surface for these age combinations. This latter observation illustrates the potential limitations of the typed point process model, because women in this age group tend to have lower HIV viral load at the population-level than men, rendering this population group of women on average less infectious than men~\citep{grabowski2017hiv, rodger2019risk}. Within the proposed Bayesian framework, such information could straightforwardly be included by using more informative priors than we have imposed here, which may play an important role for future applications of the typed point process model.

\section{Discussion}\label{sec: conclusion}
In this paper, we develop a hierarchical typed point process to learn disease transmission flows from phylogenetically reconstructed transmission pair data. 
The model is developed alongside a computationally efficient Bayesian sampler targeting the exact likelihood of the specified model. We demonstrate that our novel approach can probabilistically learn the unobserved event types --- whether or not there is transmission between a pair and in which direction the transmission occurs --- despite the large number of additional latent parameters and partially informative data. Its ability to do so is in large part owing to reformulating the inference problem in terms of a continuous spatial process. Unlike previous work~\citep{xi2021inferring}, this does not require discretization of the feature space and thus avoids keeping track of all cells in a large transmission flow matrix in computation; our method only needs to track all the data points. This advantage by construction renders our approach much more computationally efficient. In simulations and a case study on population-level HIV deep-sequence data from the Rakai Community Cohort Study in Southern Uganda, we demonstrate that the typed point process model allows inclusion of more data in analysis while accounting for the intrinsic uncertainties associated with outcomes of deep-sequence phylogenetic analyses. The primary epidemiological conclusions under the typed point process model on the Rakai data set are consistent with those drawn under the previously developed model using a subset of the data. 
We do note that among transmission events identified by our proposed approach, the odds ratio between male sources and female sources is approximately 1.32, 
which is higher than the estimate from the subset analysis with fixed event types. Most phylodynamic analyses into the age-specific drivers of HIV transmission are limited to analyses of relatively coarse age bands --- either for technical or computational reasons~\citep{de2017transmission, le2019hiv,bbosa2020phylogenetic} --- and in this context, the typed point process model extends the statistical toolkit to address epidemiologically important research questions.

Our continuous spatial process approach in modelling disease transmissions marks a contrast to the body of work that relies on count data in discretized spatial areas~\citep{berke2004exploratory,best2005comparison,wakefield2007disease,gschlossl2008modelling,mohebbi2014disease,bauer2016bayesian,johnson2019spatially}. These discrete or areal spatial models often involve the use of Gaussian Markov random fields~\citep{rue2005gaussian} or other Gaussian-based models to handle spatial dependence structures. Due to the high computational cost associated with Gaussian covariance matrix operations, inference typically entails numerical approximation techniques such as integrated nested Laplace approximations (INLA)~\citep{rue2009approximate} and still demands intensive computation. These computational challenges are inevitable when only aggregate data are available~\citep{gschlossl2008modelling,mohebbi2008geographical}, but given access to point-level data, formulating a continuous spatial model becomes more appealing than a discretized one. Directly modeling a continuous spatial process avoids expensive spatial smoothing techniques, and moreover can induce a discrete spatial model at any desired resolution, avoiding the need for manual discretization before analysis~\citep{van2017efficient, xi2021inferring}. In the context of our application, where age is known for every surveyed individual and can be treated as a continuous variable, employing a continuous spatial model provides a more natural and efficient approach compared to prior work, and in addition enables inference of the unknown event types that are associated with each phylogenetically reconstructed data point.

There are a number of future directions that one may take based on our proposed framework. It may be of interest to incorporate individual-level covariates into the spatial process, either as additional marks or latent effects of the Poisson process as in~\cite{hu2018bayesian}, or as additional covariates in the marks distributions. Second, here we did not explicitly model the sampling process of actual transmission events; but a similar approach as in~\cite{xi2021inferring} could be straightforwardly integrated to adjust the intensity function of the typed point process model. Third, extensions to non-normal components in the spatial density function mixture model can help relax certain assumptions entailed by a normal mixture model; for example, if similarity of transmission behavior is not necessarily dependant on spherical spatial proximity (an implicit assumption of the normal model), then other kernels such as the bivariate Beta kernel as in \cite{kottas2007bayesian} can  be considered.

\section*{Acknowledgements}

This work was partially supported by the National Science Foundation (DMS-2030355, DMS-2230074, and PIPP-2200047), the National Institutes of Health (R01 AI153044), and the Bill \& Melinda Gates Foundation (OPP1175094, OPP1084362). We thank the staff, investigators, and participants of the Rakai Community Cohort Study who made this work possible. We thank Mike West for helpful  discussions, and Xiaoyue Xi for helpful comments and data pre-processing. The findings and conclusions in this article are those of the authors and do not represent the official position of the funding agencies. 

\section*{Supporting Information}
All Supporting Information and Web Appendices referenced in the main text are available in the \texttt{supplement.pdf} document, attached at the end of this preprint. 

\section*{Data sharing}
All anonymized data and code to reproduce the analyses  in this paper are available at the  \texttt{GitHub} repository
\url{https://github.com/fanbu1995/HIV-transmission-PoissonProcess} under the GNU General Public License version 3.0. The deep-sequence phylogenies and basic individual-level data analysed during the current study are available in the Dryad repository (DOI: 10.5061/dryad.7h46hg2). HIV-1 reads are available on reasonable request through the PANGEA consortium; see \url{https://www.pangea-hiv.org}. Additional individual-level data are available on reasonable request to RHSP; see \url{https://www.rhsp.org}.


\setstretch{1.25}

\bibliographystyle{biom} 
\bibliography{ref}

\setstretch{1.5}


\newpage

\renewcommand{\figurename}{Web Figure}
\renewcommand{\tablename}{Web Table}

\makeatletter
\renewcommand{\ALG@name}{Web Algorithm}
\makeatother

\begin{center}
	\Huge{Supplementary Information}
\end{center}



\section*{Web Appendix A: Additional Details for Model and Inference}

Here we provide a summary of our MCMC sampling algorithm for Bayesian inference, as detailed in Section~\ref{sec: inference} of the main text.

\begin{algorithm}
	\caption{MCMC inference with data augmentation}
	\label{alg:MCMC-procedure-supp}
	
	\setstretch{1}
	
	\begin{algorithmic}[1]
		
		\begin{footnotesize}
			\Procedure{Inference}{}       
			\State Directly draw samples for $\gamma$ with
			\vspace*{-0.5em}
			\begin{equation*}
				\gamma \mid {\{\mathbf{x_i}\}, \{\mathbf{s_i}\}} \sim Ga(\alpha_0+N, \beta_0 + 1).
			\end{equation*}
			\State Randomly initialize parameter values $\Theta^{(0)}$ (except for $\gamma$).
			\State Randomly assign initial type labels $\{c_i^{(0)}\}$.
			\For{$t = 1:T$}
			\State (1) Sample $\boldsymbol{\mu}^{(t)}$, $\sigma_{\ell}^{2(t)}$, and $\sigma_{d}^{2(t)}$ conditional on $\{c_i^{(t-1)}\}$ and logit-transformed signals $\mathbf{x}_i$: 
			{\footnotesize
				\begin{align*}
					\mu_{\ell} \mid \sigma^{2(t-1)}_{\ell}, \{\ell_i\}, \{c_i^{(t-1)}\} &\sim N_{(0,\infty)}\left(\sum_{i:c_i\neq 0} \text{logit}(\ell_i), \sigma^{2(t-1)}_{\ell}/N_+\right);\\
					\mu_{d} \mid \sigma^{2(t-1)}_{d}, \{d_i\}, \{c_i^{(t-1)}\} &\sim N_{(0,\infty)}\left(\sum_{i:c_i=1} \text{logit}(d_i), \sigma^{2(t-1)}_{d}/N_1\right);\\
					\mu_{-d} \mid \sigma^{2(t-1)}_{d}, \{d_i\}, \{c_i^{(t-1)}\} &\sim N_{(-\infty,0)}\left(\sum_{i:c_i=-1} \text{logit}(d_i), \sigma^{2(t-1)}_{d}/N_1\right);\\
					\sigma^2_{\ell} \mid \mu_{\ell}^{(t)}, \{\ell_i\}, \{c_i^{(t-1)}\} &\sim \text{inv-Gamma}\left(\frac{\nu_0+N_+}{2}, \frac{\nu_0\sigma_0^2+\sum_{i:c_i\neq 0} (\text{logit}(\ell_i)-\mu_{\ell}^{(t)})^2}{2}\right);\\
					\sigma^2_{d} \mid \mu_{d}^{(t)},\mu_{-d}^{(t)}, \{d_i\}, \{c_i^{(t-1)}\} &\sim \text{inv-Gamma}\left(\frac{\nu_0+N_+}{2}, \frac{\nu_0\sigma_0^2+\sum_{i:c_i=1} (\text{logit}(d_i)-\mu_{d}^{(t)})^2 + \sum_{i:c_i=-1} (\text{logit}(d_i)-\mu_{-d}^{(t)})^2}{2}\right).
					\vspace*{-1em}
				\end{align*}
			}%
			\label{1d-normal-update}
			\State (2) For each data point $i$, sample $c_i^{(t)}$ from
			\begin{equation*}
				Pr(c_i = k \mid \Theta^{(t-1)}) \propto p_kf_k(\mathbf{s}_{i}) \phi_k(\mathbf{x}_i).
			\end{equation*}
			\State (3) sample each $\mathbf{p}^{(t)}$ conditional on $N_k$, the total number of points with $c_i^{(t)} = k$:
			\begin{equation*}
				\mathbf{p} \mid \{c_i^{(t)}\} \sim Dir\left( q_{-1} + N_{-1}, q_{0} + N_0, q_1+N_1\right).
			\end{equation*}
			\For{each type $k$}
			\State (4.a) Conditionally sample DP precision parameter $\alpha_k^{(t)}$ and  component weights $w_{kh}^{(t)}$, using updating steps described in \cite{ji2009spatial}. \label{Truncated DP step} 
			\State (4.b) For each data point $i$, sample a component latent indicator $z_i^{(t)}$ conditional on $w_{kh}^{(t)}$ and $\theta_{kh}^{(t-1)}, \Sigma_{kh}^{(t-1)}$ by
			\begin{equation*}
				Pr(z_i = h \mid w_{kh}^{(t)},\theta_{kh}^{(t-1)}, \Sigma_{kh}^{(t-1)},\mathbf{s}_{i}) \propto w_{kh}^{(t)} \varphi(\mathbf{s}_{i} \mid \theta_{kh}^{(t-1)}, \Sigma_{kh}^{(t-1)}).
			\end{equation*}
			\State (4.c) For each component $h$, sample $\theta_{kh}^{(t)}, \Sigma_{kh}^{(t)}$ conditional on all data points $\mathbf{s}_{i}$ with $z_i^{(t)} = h$:
			{\footnotesize
				\begin{align*}
					\theta_{kh} \mid \Sigma_{kh}^{(t-1)}, \{z_i^{(t)}\}, \{\mathbf{s_i}\} &\sim \text{BVN}\left((m_h(\Sigma_{kh}^{(t-1)})^{-1}+\Sigma_0^{-1})^{-1}\left(\sum_{i:z_i^{(t)}=h}(\Sigma_{kh}^{(t-1)})^{-1}\mathbf{s_i} + \theta_0\Sigma_0^{-1}\theta_0\right),(m_h(\Sigma_{kh}^{(t-1)})^{-1}+\Sigma_0^{-1})^{-1}\right);\\
					\Sigma_{kh} \mid \theta_{kh}^{(t)},\{z_i^{(t)}\},\{\mathbf{s_i}\} &\sim \text{inv-Wishart}\left(\nu+m_h, \left(S_0^{-1}+\sum_{i:z_i^{(t)}=h}(\mathbf{s_i}-\theta_{kh}^{(t)})(\mathbf{s_i}-\theta_{kh}^{(t)})^T\right)^{-1}\right).
				\end{align*}
			}%
			\label{2d-normal-update}
			\EndFor
			\EndFor
			\State \textbf{Return} MCMC samples for parameters $\Theta$ and type labels $\{c_i\}$
			\EndProcedure
		\end{footnotesize}
		
	\end{algorithmic}
	\setstretch{1.0}
\end{algorithm}

\section*{Web Appendix B: Supplemental Materials for Simulation Study}

\subsection*{Parameters and settings in the simulations}

\begin{figure}[ht]
	\centering
	\includegraphics[width=0.7\textwidth]{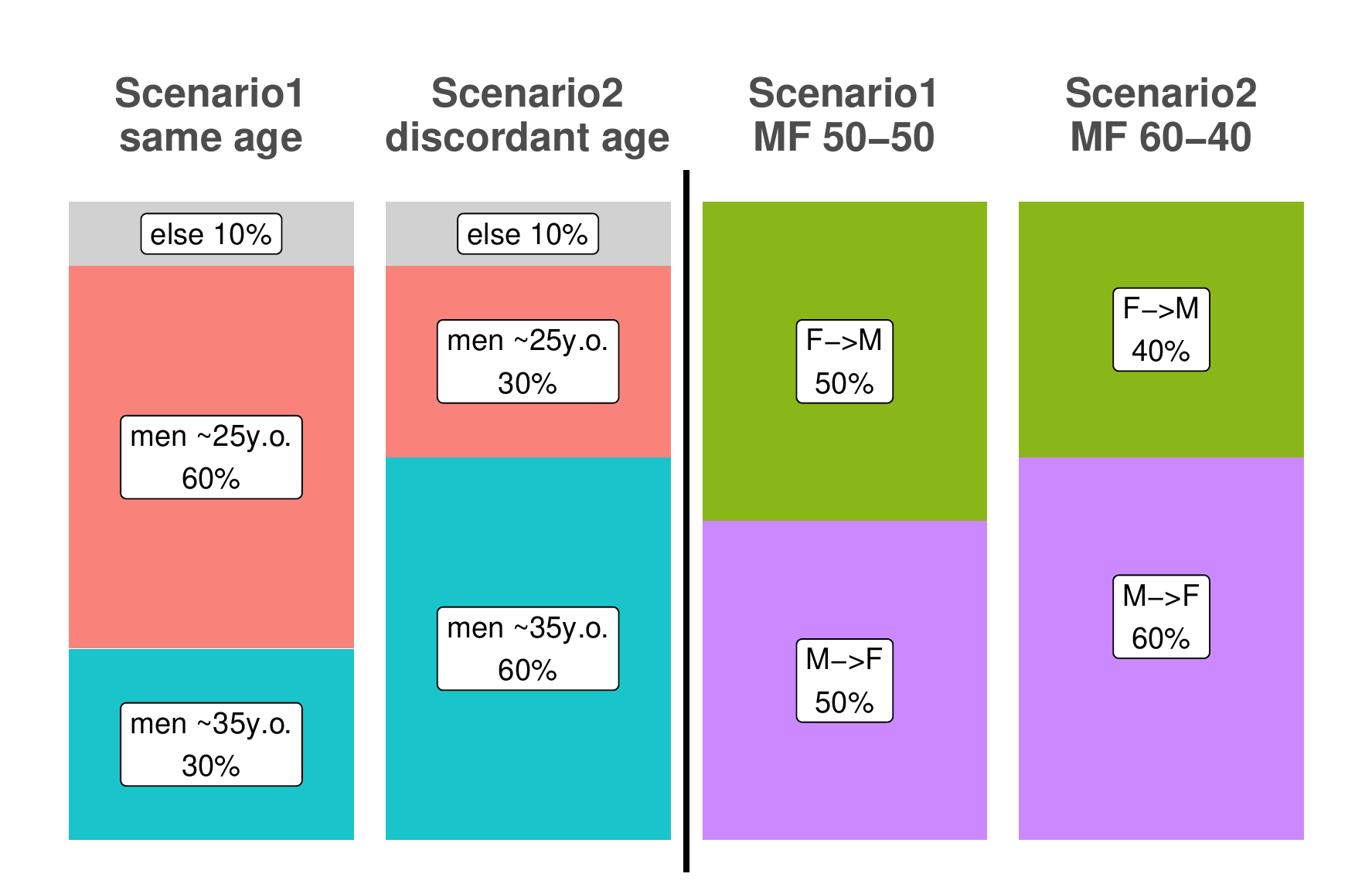}
	\caption{Graphic illustration of the simulation setup. We consider two different scenarios for each epidemiological question of interest. (1) Age of male sources for young women between 15 and 24 (left panel); scenario ``\textbf{same age}'' has most young women infections attributable to young men around 25 who are of similar age to those infected, whereas in scenario ``\textbf{discordant age}'' most such transmissions are attributable to older men around 35. (2) Proportions of male-to-female and female-to-male events; for scenario ``\textbf{MF 50-50}'' the two transmission directions have equal incidents, while in scenario ``\textbf{MF 60-40}'' there are slightly more MF transmissions which contribute to 60\% of total infections. }
	\label{fig:simulation_illustration}
\end{figure}

Below we detail the parameter choices in the simulation study described in Section~3 of the main text. We also include a diagram that illustrates the different scenarios compared in our simulations in Web Figure~\ref{fig:simulation_illustration}. 

\paragraph{1: age structure of male sources for infections in young women}

We consider two scenarios with different setup in the BVN mixture model of the spatial density function $f_1(\cdot)$ for MF transmissions:
\begin{itemize}
	\item \textbf{same age}: For women aged between 15-24, infections from younger men take the majority; here, for density function $f_{1}(\cdot)$, we set the component centered at $(35,20)^T$ (i.e., older men sourced transmission) to have mixture weight \textbf{$0.3$}, and the component centered at $(25,20)^T$ (i.e., younger men sourced transmission) to have mixture weight \textbf{$0.6$}.
	\item \textbf{discordant age}: For women aged between 15-24, infections from older men take the majority; here, for function $f_{1}(\cdot)$, we set the component centered at $(35,20)^T$ to have mixture weight \textbf{$0.6$}, and the component centered at $(25,20)^T$ to have mixture weight \textbf{$0.3$}. 
\end{itemize}

For each scenario, the other $10\%$ probability mass for MF transmissions is spread across all other bivariate normal components of the density function $f_1$.

\paragraph{2. proportions of MF and FM transmission events}

We consider the following two scenarios with different value choices for the type probability $\mathbf{p}$:

\begin{itemize}
	\item \textbf{MF 50-50}: There are equal proportions of MF and FM events, and each contribute to about 50\% of all real transmission events; we set 25\% of events to be non-transmission events, and thus we have $p_0 = 0.25, p_{-1} = p_{1} = 0.375$. 
	\item \textbf{MF 60-40}: There are more MF events than FM events, and MF events contribute to about 60\% of all transmissions; again, we set 25\% of events to be non-transmissions, and thus we have $p_0 = 0.25, p_{-1} = 0.3, p_{1} = 0.45$. 
\end{itemize}

To simulate the spatial patterns, we use $6$ different BVN mixture components to construct the three density functions $f_{0}, f_{-1}, f_{1}$; for the signal distributions, we set $\mu_{\ell} = 2$, $\mu_{d} = 1.5, \mu_{-d} = -1.5$, and $\sigma_{\ell}^2 = \sigma_{d}^2 = 1$. 

For each simulation, we run the Bayesian MCMC sampler for $3000$ iterations with a $1000$-iteration burn-in period, and adopt the following hyper-parameters for the priors: $a_0 = 1, b_0 = 0.02$ (prior for $\gamma$), $\nu_0 = 2, \sigma_0^2 = 1$ (priors for $\sigma_{\ell}^2$ and $\sigma_{d}^2$), $q_{i}=1, i=1,2,3$ (prior for $\mathbf{p}$), $\theta_0 = (0,0)^T, \Sigma_0 = \bigl( \begin{smallmatrix}10^4 & 0\\ 0 & 10^4\end{smallmatrix}\bigr)$ (priors for $\theta_{kh}$'s), $\nu = 2, S_0 = I_2$ (priors for $\Sigma_{kh}$'s), and $a=2,b=3$ (priors for $\alpha_k$'s). We have experimented with various hyper-parameter values and have found that the results are not sensitive to value changes within reasonable ranges. 

\subsection*{Simulation study results}



\begin{figure}[ht]
	\centering
	\hspace*{-0.25in}
	\includegraphics[width = \textwidth]{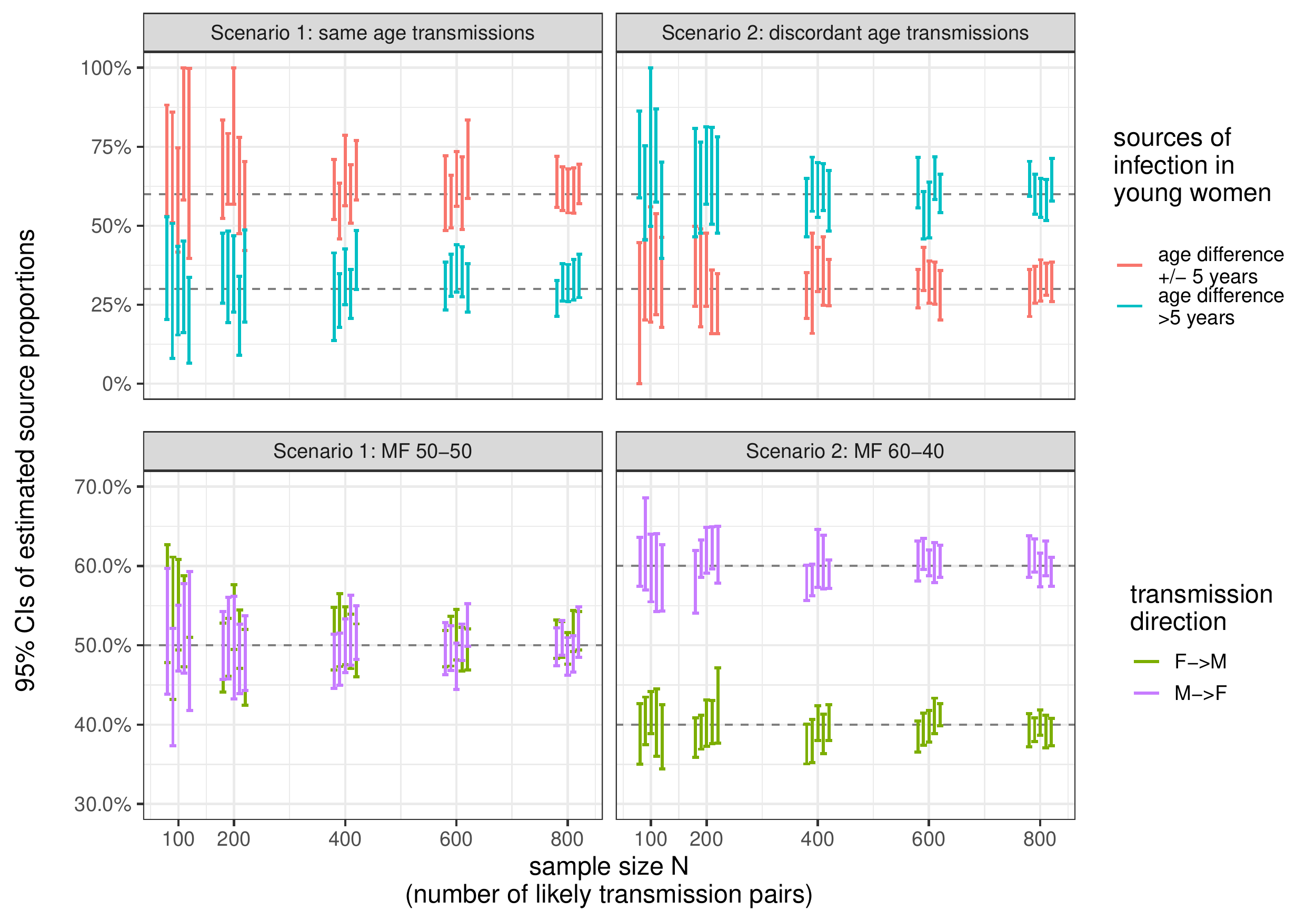}
	\caption{95\% posterior credible intervals of the simulation study. Results are shown for \textbf{five} (5) randomly selected simulation run for each $N$ and each scenario. 
		Each error bar shows the 95\% posterior credible interval.
		The true parameter values are marked in dashed lines in each subplot. }
	\label{fig:CIs_by_N_all_views}
\end{figure}

In addition to the posterior medians results shown in the main text, within each simulation run, we can also look at the posterior credible intervals. 
In Web Figure~\ref{fig:CIs_by_N_all_views}, we plot the 95\% posterior credible intervals for the proportions of male source age groups in each scenario in the top row, and for proportions of each transmission directions as well in the bottom row. 
For each sample size $N$ and scenario setting, the plotted credible intervals are acquired from \textbf{five} (5) randomly selected simulations among $100$ total runs. 
We can see that the majority of these credible intervals do cover the truth (marked by dashed lines in each sub-plot), and when $N$ gets larger, the credible intervals get narrower, indicating less uncertainty when more data are available. 

\begin{figure}[ht]
	\centering
	\begin{subfigure}[t]{0.48\textwidth}
		\centering
		\includegraphics[width=\textwidth]{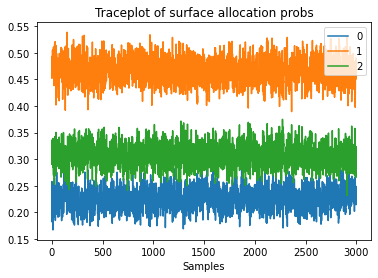}
		\caption{Traceplot of the surface probabilities (entries of $\mathbf{p}$). $0=$ non-event, $1=$ MF transmission surface, $2 =$ FM transmission surface.}
		\label{fig:probs-traceplot}
	\end{subfigure}
	\hfill
	\begin{subfigure}[t]{0.48\textwidth}
		\centering
		\includegraphics[width=\textwidth]{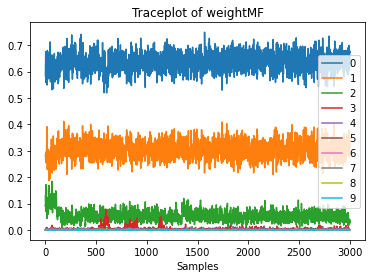}
		\caption{Traceplot of the BVN component mixture weights (with label switching accounted for) for the male-to-female (MF) transmission surface. Each number/color represents a unique BVN component. }
		\label{fig:weightMF-traceplot}
	\end{subfigure}
	\hfill
	\caption{Traceplots for transmission surface type probabilities (subplot (a)) and BVN mixture component weights for FM surface (subplot (b)) throughout the 3000 iterations of the MCMC sampler. Example from a randomly picked simulation run with $N=600$.}
	\label{fig:sim-traceplots}
\end{figure}

Moreover, within each run of the simulation study, we can also inspect the MCMC traceplots (such as the ones shown in Web Figure~\ref{fig:sim-traceplots}) to check for convergence and inference quality, which is also similarly done for the real data case study. 
For example, in subplot (a), we plot the values of entries in $\mathbf{p}$ (the probability/probability vector for the three transmission types) sampled across all 3000 iterations of the MCMC sampler, where the three lines represent MF, FM and non-event transmission surfaces from top to bottom. 
Inspecting this traceplot, we can see that the sampled probability values stabilize after the initial 500 iterations or so (used as the ``burn-in'' period) and, in later iterations, fluctuate around a value close to ground truth (in this case, MF and FM probabilities have an approximate ratio of 60\% versus 40\%), which is an indicator of convergence. 
Also, in subplot (b), we show the traceplot of the sampled mixture weights (of the BVN components) for the MF surface. 
Again, we see that the weights have stabilized in later iterations after some exploration in the early steps, and there are three dominating components where the two biggest ones have weights around 0.6 and 0.3, respectively, which is very close to the true mixture weight values.  
Note that in real data analysis, we can also use the traceplots as a graphical diagnostic tool to check MCMC convergence.

\section*{Web Appendix C: Supplemental Materials for the Case Study}

\subsection*{Full analysis with flexible point types}

In the full analysis, we pre-specify$\mu_{d} = 1.5$ and $\mu_{-d} = -1.5$ to imply that the $d_i$'s with $i=1$ are centered around $0.817$ and the $d_i$'s with $i=-1$ are centered around $0.182$. 
We note that analysis results are not considerably sensitive to the choices of these parameters, and through experiments, we've found that altering these values within a reasonable range produces consistent results.  

Moreover, for a pair with $d_i=1$ or $0$ (extremely large or small direction scores), in the inference process, we conditionally assign it to the MF or FM surface depending on its $\ell_i$ score - if $c_i$ is sampled to be nonzero in the inference algorithm, then we would set $c_i=1$ if $d_i=1$ and otherwise set $c_i=-1$.

\subsection*{Additional analysis results}

Here we provide a graphical illustration as supplement to Table 1 in the main text. 
As described in Section 5, our model (``\textbf{model}'') recognized more data points as real transmission events compared to the fixed-type analysis (``\textbf{fixed}''), thus effectively leveraging more information from the data. 

\begin{figure}[ht]
	\centering
	\includegraphics[width = 0.8\textwidth]{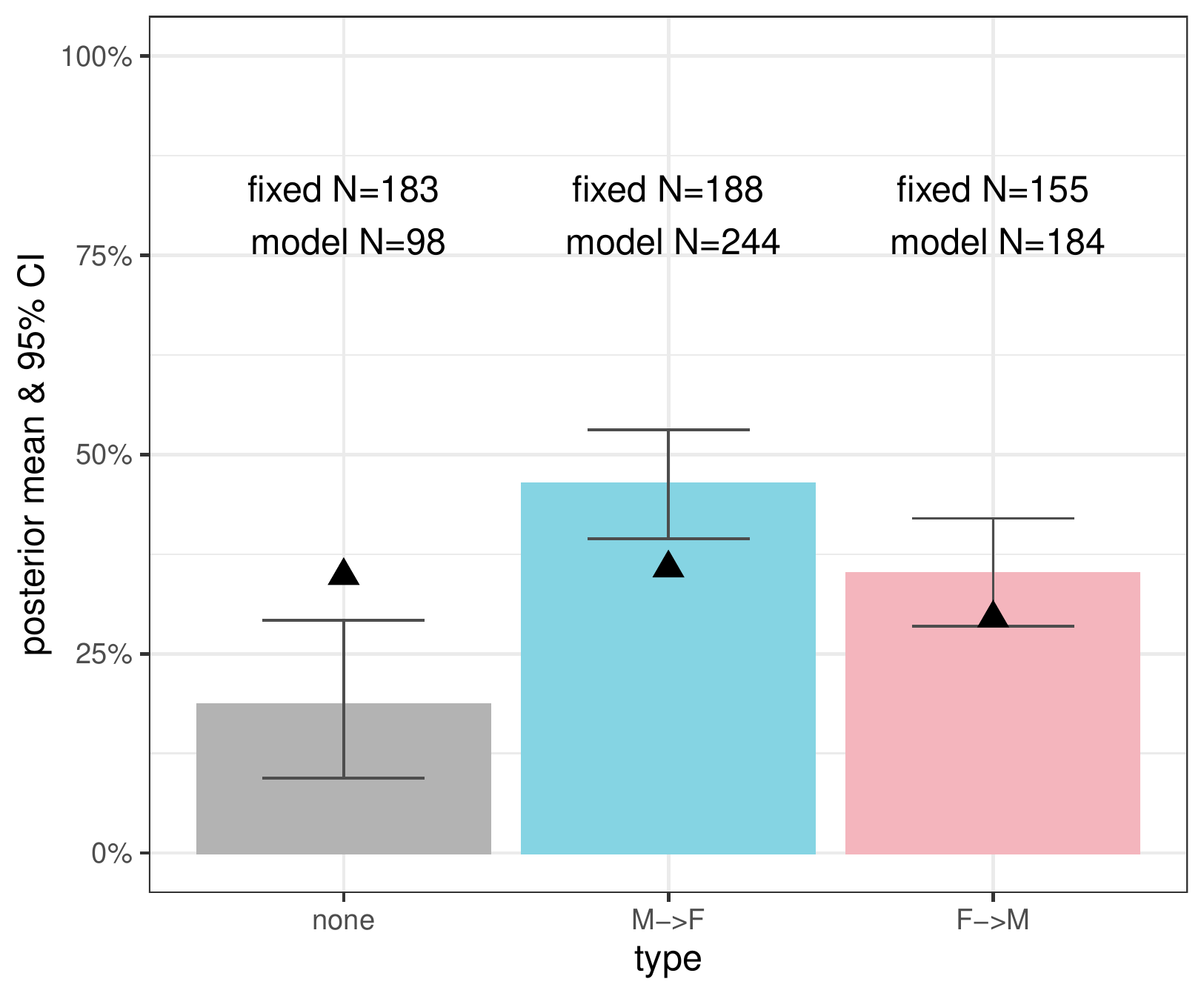}
	\caption{Proportions of data points assigned to different type labels (MF, FM or none) identified by the full model (colored bars), compared to partial analysis with fixed point types (marked by \mytriangle{black}'s). The colored bars show the posterior median proportions of type labels, with black errorbars showing the 95\% credible intervals; the number of points assigned to each type (using posterior median numbers for the full model) is also annotated on the plot. The full model consistently identifies significantly more data points as MF or FM transmission events than the fixed model, while the relative ratios between MF and FM event counts are similar between the full and partial analyses (fixed: $\text{MF}/\text{FM} \approx 1.22$, full model: $\text{MF}/\text{FM} \approx 1.32$).}
	\label{fig:type-proportions}
\end{figure}

\subsection*{Source age distribution by specific recipient age groups}

\begin{figure}[h]
	\centering
	\includegraphics[width=0.48\textwidth, page=3]{figures/source_freq_specTreat_Jan2022_titled.pdf}
	\includegraphics[width=0.48\textwidth, page=4]{figures/source_freq_specTreat_Jan2022_titled.pdf}
	~
	\includegraphics[width=0.48\textwidth, page=1]{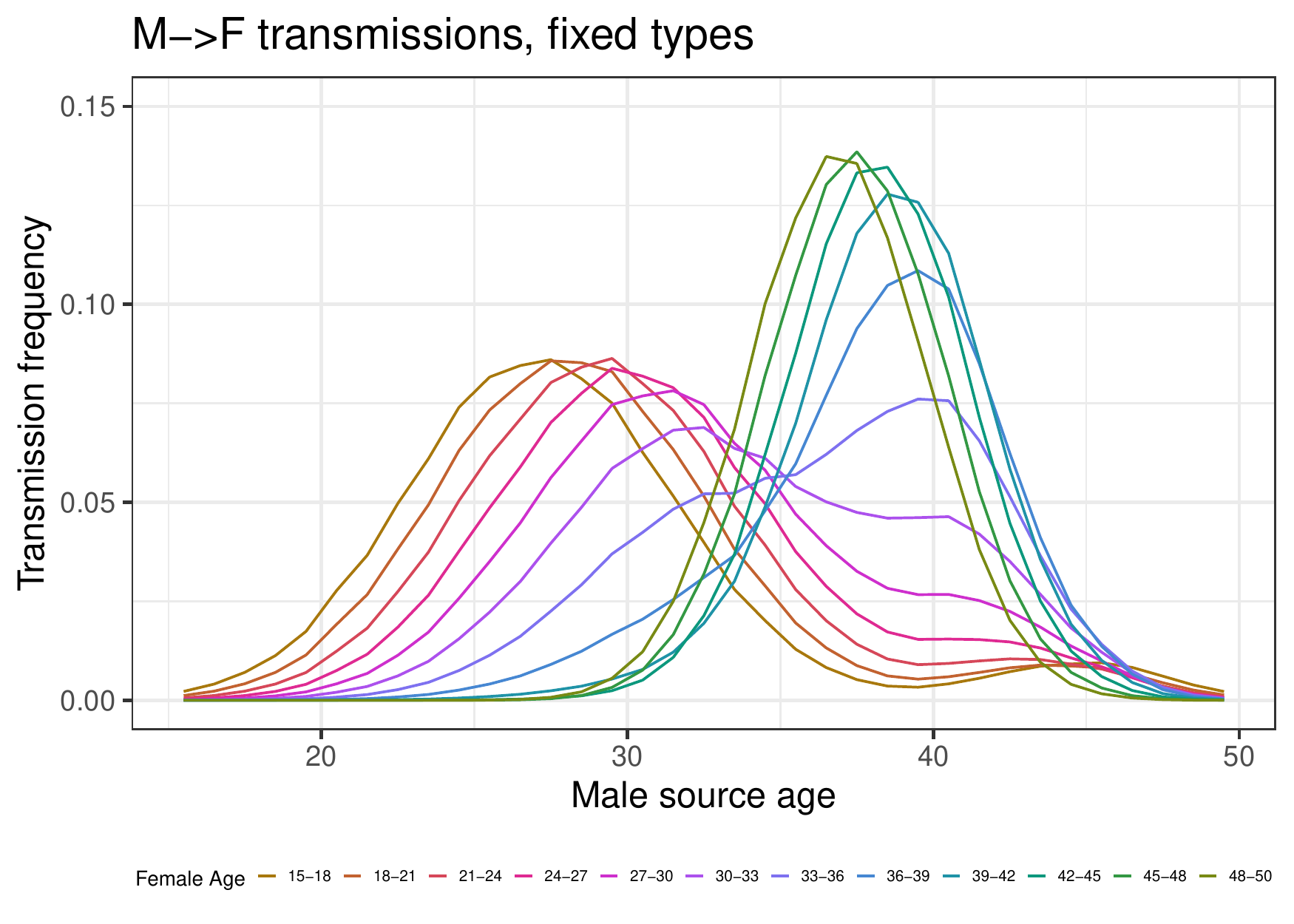}
	\includegraphics[width=0.48\textwidth, page=4]{figures/source_freq_fixedAlloc_Jan2022_titled.pdf}
	
	\caption{Source age distribution for recipients in each 3-year age band. 
		\textbf{Top row}: results learned from full analysis without pre-fixing point type labels.
		\textbf{Bottom row}: results learned from the partial analysis with pre-fixed point types.
		Left column: male sources and female recipients; right column: female sources and male recipients. }
	\label{fig:source-prob-by-age}
\end{figure}

In Web Figure~\ref{fig:source-prob-by-age}, for each 3-year age band of recipients, we plot the relative frequency curve for the age of heterosexual sources.
On the left column, we show inferred results for male to female (MF) transmissions -- each curve corresponds to each 3-year age group of female recipients (e.g., the leftmost curve represents the age distribution of male sources for women between 15 to 18); 
on the right column, we show results in the same manner for female to male (FM) transmissions. 
The full analysis (flexible point types) results are presented in the top row, while partial analysis (fixed point types) results are shown on the bottom. 

We can see that for different age groups of recipients, the age distributions of their sources could be very different. 
On the left column (male-to-female transmissions), we can see that younger women may get infected by both younger and older men, but older women main get infected by similarly aged men. 
On the right side (female-to-male transmissions), we see a more age assortative behavior (young men mainly get infected by young women and older men get infected by older women); however, we also see a bump of female sources at around age 40 for younger men, which also agrees with the general findings discussed in the main paper.

\subsection*{How our model leverages uncertain data points}
\label{sec: case-study-flexible-points-showcase}

A central feature of our proposed model is to probabilistically and flexibly classify point types, particularly for the ones with medium direction scores and/or low linkage scores (the uncertain points). 
Those uncertain points play an important role in inferring the spatial patterns, borrowing information across different types, and helping leverage spatial information for point type allocation. 

For a data point $i$, the posterior mean probability vector $\hat p_i = (\hat p_{i,0},\hat p_{i,1},\hat p_{i,-1})^T$ for its type indicator $c_i$ represents the level of uncertainty we have about its type. (Here, for example, $\hat p_{i,1}$ is the relative frequency of $i$ being assigned to the MF surface across all posterior samples. ) \emph{If the entropy of $\hat p_i$ is high, then we have high uncertainty about $i$'s type.} 

\begin{figure}[ht]
	\centering
	\includegraphics[width=0.48\textwidth]{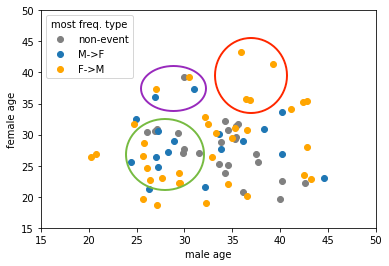}
	\includegraphics[width=0.48\textwidth]{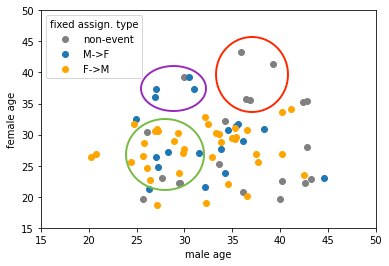}
	\caption{Flexible-type data points with classification entropy $> 0.8$ (77 points in total). Left: point color represents the most frequent type assignment; right: point color represents the fixed threshold classification. }
	\label{fig:flexible-points-showcase}
\end{figure}

In Web Figure~\ref{fig:flexible-points-showcase}, we plot the 77 data points with classification entropy $> 0.8$ (as a reference, $(0.1, 0.2, 0.7)^T$ would have entropy $= 0.802$). On the left, the points are colored according to the types they are \emph{most frequently} assigned to in the posterior samples (as in the sampling iterations of the full model).
On the right, the color represents the type a point would get assigned to given fixed pre-classification (as in the flexbile-type analysis). 

One of the most distinct differences between the two plots is the classification of the three points in the upper right corner (inside the red circle). 
In the full model, they tend to be assigned to the FM surface as their linkage scores are relatively high (close to the $0.6$ fixed threshold) and their locations are close to the FM surface mass near $(40, 40)$. 
Therefore, these points contribute (at least probabilistically) to characterizing the transmission flow patterns from older women to older men. 

Another notable difference is within the green circles of the two plots. 
The full model would, in fact, switch the type allocations for most of those points across the iterations of the inference algorithm, since they are mostly flexible-type points with close-to-threshold direction and linkage scores. 
(Note that this region is a high-density region for both MF and FM surfaces.)
By exploring the possible configurations of data point classification, the model effectively allows spatial information to be shared between different surfaces (especially between MF and FM transmission surfaces) while accounting for the uncertainty in the point types.
This is exactly what we hope to achieve through a hierarchical model.

Finally, within the purple circles, the two points at the upper left corner tend to be more frequently classified as FM transmission events, as opposed to the MF classification using fixed thresholds. 
This is actually in part due to the slightly different spatial patterns of the two transmission directions learned from all other data points: 
in the FM surface (bottom-middle panel in Figure~5 of the main text), there is a slightly denser mass for older-women to younger-men transmissions, and hence these two points tend to be assigned to the FM surface slightly more frequently.
This is an example of how the learned spatial patterns help inform data point classification throughout the iterations of the inference procedure, thanks to joint modeling of the spatial process and the signal distributions as described in Section 2 of the main text.


\end{document}